\documentclass[aps,
showpacs,
twocolumn,
nofootinbib,
nobibnotes,
superscriptaddress,
longbibliography,
amsmath,amssymb,prd]{revtex4-2}

\usepackage[skins,theorems]{tcolorbox}
\usepackage{color}
\usepackage{soul}
\usepackage{centernot}
\usepackage{graphicx}
\graphicspath{{./Figures/}} 
\usepackage{dcolumn}
\usepackage{siunitx}
\usepackage{bm}
\usepackage{bbold}
\usepackage{braket}
\usepackage{esdiff}
\usepackage{hyperref}
\usepackage[mathlines]{lineno}
\usepackage{subfigure}

\begin{document}
\title{Observer dependence of photon bunching:
\\
The influence of the relativistic redshift on Hong-Ou-Mandel interference}

\author{Roy Barzel}
\affiliation{ZARM, University of Bremen, 28359 Bremen, Germany}
\author{David Edward Bruschi}
\affiliation{Institute for Quantum Computing Analytics (PGI-12), Forschungszentrum J\"ulich, 52425 J\"ulich, Germany}
\author{Andreas W. Schell}
\affiliation{Institut for Festkörperphysik, Leibniz Universität Hannover, Appelstr. 2, 30167 Hannover, Germany }
\affiliation{ Physikalisch-Technische Bundesanstalt, Bundesallee 100, 38116 Braunschweig, Germany }
\author{Claus L\"ammerzahl}
\affiliation{ZARM, University of Bremen, 28359 Bremen, Germany}
\affiliation{DLR-Institute for Satellite Geodesy and Inertial Sensing,
c/o University of Bremen, Am Fallturm 2, 28359 Bremen, Germany}
\affiliation{Institute of Physics, Carl von Ossietzky University Oldenburg, 26111 Oldenburg, Germany}


\begin{abstract} 
We study the influence of the relativistic redshift on Hong-Ou-Mandel (HOM) interference, and present a genuine quantum test of general relativity. We use Glauber's theory of quantum coherence to predict the coincidence probability of realistic broadband photons in HOM-experiments in a non-relativistic setting. We extend the quantum field theoretical framework previously developed to describe the deformation of the spectral profile of single photons in curved spacetimes to a multi-photon framework, which is exact for inertial observers in a flat spacetime and an approximation when observers are located in a curved spacetime. We find that, in case of frequency entangled photons, a mutual redshift between the sender and the receiver can change the coincidence statistics from photon bunching to photon anti-bunching, and vice versa. This implies that the (anti-) symmetry of the photonic spectral wave function is an observer dependent notion, and that this can be probed via HOM-experiments in a relativistic setting.
\end{abstract} 

\maketitle



\section{\label{sec:introduction} Introduction}
Exploring the interplay between general relativity and quantum mechanics is one of the major goals of modern physics \citep{feynman2018feynman}.  A central problem in the development of a theory that describes the behavior of quantum particles in curved spacetimes is the lack of experimental evidence providing guidance. This is mostly rooted in the huge discrepancy of the parameter regimes where the two theories become relevant. While relativistic effects typically occur at macroscopic scales and large energies, quantum effects usually manifest at the microscopic scale, where few elementary objects interact. However, in quantum theory there is no length scale that restricts its applicability. Moreover, in recent decades, fuelled by advances in experimental capabilities and control, several experiments have been conducted to reveal the effects of gravity on quantum systems. For example, the non-relativistic gravitational phase of neutrons was measured in the famous Colella, Overhauser and Werner (COW) experiment \citep{COW}, and the general relativistic redshift on the frequency of single photons \citep{PhotRedshift,PhotRedshift2} or of clocks located on Earth or moving \cite{Mueller:Peters:2010}, was detected. These early successes indicate that it is possible to investigate the overlap of relativity and quantum mechanics within accessible energy scales and environments, and that the limit is mostly technological.

The rapid technological progress experienced in the past decades offers the opportunity to propose experiments that operate in parameter regimes where both general relativity and quantum mechanics play a non-negligible role \citep{SecondQuantumRevolution}. For example, the recent demonstration of entanglement distribution over more than $1200$km \citep{1200} opened the way for many long-distance quantum applications, such as the first satellite-to-ground quantum key distribution (QKD) implementation by the Micius satellite of the Chinese Academy of Sciences \citep{liao2017satellite,ren2017ground,yin2017satellite}, and intercontinental quantum-secured data transfer \citep{liao2017space,liao2018satellite}. Proposed  quantum technologies often rely on the use of genuine quantum features, such as entanglement, shared between several distant systems or particles. Quantum mechanics has so far been successfully employed as the only formal framework to describe these scenarios. However, it is reasonable to assume that relativistic effects cannot be neglected anymore when correlated quantum systems are distributed over large distances through the inhomogeneous gravitational field of the Earth. 

Relativistic and quantum information aims at understanding how gravity affects quantum information tasks and protocols \cite{Mann:Ralph:2012}. One  avenue has demonstrated that gravity affects the quantum state of realistic broadband photons propagating in curved spacetime \citep{Bruschi1,Bruschi2,Bruschi4}. It was shown that the effects can also be employed for sensing \cite{Bruschi2}. Furthermore, quantum correlations, such as squeezing and entanglement, have been shown to potentially enhance the sensitivity of interferometers in this context \citep{schnabel2017squeezed}, and to improve the accuracy in clock synchronization \citep{giovannetti2001quantum,giovannetti2004quantum}. The latter is needed in the global positioning system (GPS) or high precision metrology in general. 
It has become evident that the development of modern technologies to be deployed in contexts where gravity plays a significant role, such as a GPS with quantum systems as atomic clocks, requires a full characterization of the properties of quantum systems in curved spacetime. 

Indistinguishability of physical systems is one of the important defining features of quantum systems without classical analogue. In the case of bosons, indistinguishability leads to intriguing phenomena, such as photon-bunching within interference experiments, first witnessed in the pioneering Hong-Ou-Mandel (HOM) experiment \citep{HOM}. Only recently it was shown that HOM-interference bears the potential to enhance the accuracy of clock synchronization \citep{quan2016demonstration,valencia2004distant}, as already predicted theoretically \citep{giovannetti2001quantum}. Consequently, HOM-like schemes are candidates for space-based implementations where indistinguishable correlated photons are exchanged over intercontinental distances. Relativistic redshift, which plays a role already in ordinary GPS, is expected to play an important role in these scenarios as well. We conclude that HOM-interference can constitute one of the most promising routes to design genuine quantum tests of general relativity. It is therefore important to understand the effect of relativity on HOM-interference-based implementations. 

In the present paper we investigate the influence of relativistic redshift on HOM-interference experiments. 
This has a fundamental and a practical aspect. On the fundamental side, it is of interest to reveal which scenarios witness signatures of general relativistic effects that can be observed in HOM-experiments. On the practical side, it is important to find the regimes where relativity affects quantum technological applications, and to quantify the magnitude of the effects. 
In particular, we want to quantify the signatures of general relativistic effects on observable HOM-interference patterns, which in turn will inform us on potential new challenges to be faced in the development of novel quantum technological applications. We employ the quantum field theoretical framework developed to quantify the influence of the relativistic redshift on single photons \citep{Bruschi1,Bruschi2}, and extend this body of work to the case of two (or more) photons. This, in turn, allows us to quantify relativistic effects on two-photon HOM-interference. Our approach can be applied to the case of inertial observers in a flat spacetime, as well as the case of scalar optics in (weakly) curved spacetime, which is the regime considered in our analysis. We also provide a comparison between the effects obtained when employing pairs of frequency uncorrelated photons, frequency detuned photons and frequency-entangled photons. Finally, we derive the experimental conditions necessary to detect the effects of relativistic redshift on the HOM-interference pattern with frequency entangled photons. 

This work is organized as follows. In Section~\ref{sec:HOM} we briefly review our results on HOM-interference from previous work \citep{Barzel2}. In Section~\ref{sec:REL} we provide the tools to compute the relativistic deformation of the spectral wave function. In Section~\ref{sec:RELHOM} we compute the effects using the newly developed methods. In Sections~\ref{sec:outlook} and \ref{sec:conclusions} we provide outlook and conclusions for our work.

\section{\label{sec:HOM} Hong-Ou-Mandel Interference}
\begin{figure}[h]\centering
  \includegraphics[width=0.75\linewidth]{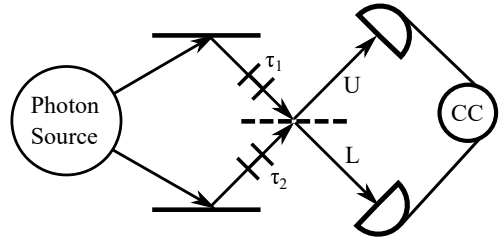}
  \caption{Scheme of Hong-Ou-Mandel interference. Two photons are subject to optical delays $\tau_{1,2}$ and interfere on a common beam splitter to produce a joint detection statistic at the detectors $U$ and $L$, which may be evaluated with a coincidence count (CC) logic.}
\label{fig:HOM}
\end{figure}

In this section we briefly review a theoretical framework previously developed to predict the detection statistics of HOM experiments with realistic broadband photons. We leave details to the interested reader \citep{Barzel2}. In this section we work in a non-relativistic context, i.e., all temporal quantities are functions of a universal background time and not of the proper time of the observers. The basic scheme of HOM-interference is shown in Figure \ref{fig:HOM}. 

\subsection{Detection statistics}
A general two-photon state $\ket{\psi(\tau_1,\tau_2)}$ in the context of HOM-interference is given by 
\begin{align}
\ket{\psi(\tau_1,\tau_2)}=\int d \bm{\omega}_1 d \bm{\omega}_2\, \Phi(\bm{\omega}_1,\bm{\omega}_2)e^{i\omega_1\tau_1}e^{i\omega_2\tau_2}
\label{EQ:2phot}
\hat{a}_{\bm{\omega}_1}^\dagger \hat{a}_{\bm{\omega}_2}^\dagger \ket{0},
\end{align}
where the bold font $\bm{\omega}$ denotes the set of parameters that characterize all degrees of freedom (DOFs) of a single photon and the $\hat{a}^\dagger_{\bm{\omega}}$ are bosonic creation operators. We treat the frequency DOF $\omega$ separately from the other photonic DOF $\bm{\sigma}$ and write $\bm{\omega}=\{\omega,\bm{\sigma} \}$, thereby defining the alternative notation of the photonic wave function $\Phi(\bm{\omega}_1,\bm{\omega}_2)=\Phi(\omega_1,\bm{\sigma}_1,\omega_2,\bm{\sigma}_2)=\Phi_{\bm{\sigma}_1\bm{\sigma}_2}({\omega}_1,{\omega}_2)$.

The (unnormalized) probabilities ${P}_{\bm{\sigma}_1\bm{\sigma}_2}(\tau_1,\tau_2)$ to detect one photon in a quantum state, which is characterized by $\bm{\sigma}_1$ and another photon in a quantum state, which is characterized by $\bm{\sigma}_2$, in the two-photon quantum state (\ref{EQ:2phot}) have been obtained previously \citep{Barzel2}, and read
\begin{align}
{P}_{\bm{\sigma}_1\bm{\sigma}_2}(\tau_1,\tau_2)=&\int d \omega_1d \omega_2\ \left[|\Phi_{\bm{\sigma}_1\bm{\sigma}_2}(\omega_1,\omega_2)|^2\right.
\nonumber
\\
&\hspace*{20mm} +\left.|\Phi_{\bm{\sigma}_2\bm{\sigma}_1}(\omega_2,\omega_1)|^2\right]
\nonumber
\\
&\hspace*{-20mm} +2\Re\left\{\int d \omega_1d \omega_2 \Phi_{\bm{\sigma}_1\bm{\sigma}_2}(\omega_1,\omega_2)\Phi^*_{\bm{\sigma}_2\bm{\sigma}_1}(\omega_2,\omega_1)\right.
\nonumber
\\
&\times\left. e^{-i(\omega_1-\omega_2)(\tau_1-\tau_2)}\right\}.\label{EQ:P}
\end{align}
We also call the probabilities ${P}_{\bm{\sigma}_1\bm{\sigma}_2}(\tau_1,\tau_2)$ in (\ref{EQ:P}) the \textit{detection statistics} of a HOM-experiment. Here, $\tau_1$ and $\tau_2$ are optical delays that can be applied independently to wave packets in HOM-interferometry \citep{pradana2019quantum}. Note that the probabilities are always a function of the difference $\Delta\tau:=\tau_1-\tau_2$ as can be seen from (\ref{EQ:P}). Nevertheless we keep track of both delays $\tau_1$ and $\tau_2$ in our work since this will be necessary when considering relativistic effects on the detection statistics in Section~\ref{sec:RELHOM}.

\subsection{Two-photon sources}\label{subsec:HOM}
In this section apart from the frequency DOF we consider the spatial DOF of the photons (i.e. $\bm{\sigma}=(U,L)$) as the only left photonic DOF and do not consider for instance the polarization of the photons and so forth. We consider two-photon sources which posses photonic wave functions which can be written in matrix form (see \citep{Barzel2} for details) as
\begin{align}
\Phi_{\bm{\sigma}_1\bm{\sigma}_2}({\omega}_1,{\omega}_2)=\phi(\omega_1,\omega_2)
\begin{pmatrix}
e^{i\theta} & +1 \\
-1 & -e^{-i\theta}
\end{pmatrix},\label{EQ:Pmat}
\end{align}
where $\theta$ is the phase associated with BS reflection. The row and column numbering of the matrix in (\ref{EQ:Pmat}) is $U,L$ and $\phi(\omega_1,\omega_2)$ is the joint photonic \textit{spectral wave function}, for which we may impose w.l.o.g. the normalization condition $\int d\omega_1 d\omega_2 |\phi(\omega_1,\omega_2)|^2=1$.

Inserting (\ref{EQ:Pmat}) into (\ref{EQ:P}) and computing the coincidence detection probability $P^c=P_{UL}+P_{LU}$, i.e. the probability to detect one photon at one detector $U$ or $L$ and the other photon at the other detector $L$ or $U$ yields after subsequent normalization (such that $P_{UU}+P_{LL}+P_{UL}+P_{LU}=1$) the result
\begin{align}
P^c(\tau_1,\tau_2)=\frac{1}{2}[1-d_{\phi}(\tau_1,\tau_2)],\label{EQ:Pc}
\end{align}
where
\begin{align}
d_\phi(\tau_1,\tau_2)=2\Re\left\{\int d\omega_1 d\omega_2 \phi(\omega_1,\omega_2)\phi^*(\omega_2,\omega_1)e^{-i\Delta\omega\Delta\tau}\right\}\label{EQ:d}
\end{align}
with $\Delta\omega=\omega_1-\omega_2$ and $\Delta\tau=\tau_1-\tau_2$ contains the dependence on the optical delays $\tau_1$ and $\tau_2$ and depends on the spectral profile $\phi(\omega_1,\omega_2)$ of the two photons.

In section \ref{sec:RELHOM} we investigate the influence of relativistic effects on HOM-experiments which are operated with three different types of two-photon sources each owning a different two-photon spectral profile. We investigate a source of spectrally indistinguishable frequency uncorrelated photons (employed in the original work of Hong, Ou and Mandel \citep{HOM}) which features a spectral profile
\begin{align}
{\phi_\mathrm{HOM}}(\omega_1,\omega_2)=f_{\mu}(\omega_1)f_{\mu}(\omega_2),\label{EQ:GG}
\end{align}
with
\begin{align}
f_{\mu}(\omega)=\frac{1}{\sqrt{\sqrt{2\pi}\xi}}e^{-\frac{\left(\omega-\mu\right)^2}{4\xi^2}}.\label{EQ:Gauss}
\end{align} 

Further we investigate parametric down converting photon sources generating spectrally distinguishable frequency detuned photons with a spectral profile of
{\small
\begin{align}
{\phi_\mathrm{f.d.}}(\omega_1,\omega_2)=\sqrt{\frac{2}{\pi\xi}}\delta(\omega_p-\omega_1-\omega_2)\,\mathrm{sinc}\left(\frac{\omega_1-\omega_2-\mu}{\xi}\right),\label{EQ:FE1}
\end{align}
}
where $\omega_p$ is the pump frequency of the down converting process and $\mu$ is the frequency separation/detuning of the photons.

Lastly we investigate sources of spectrally indistinguishable frequency entangled photons owning the spectral profile
\begin{align}
\phi_\mathrm{f.e.}(\omega_1,\omega_2)=\mathcal{N} [{\phi_\mathrm{f.d.}}({\omega}_1,{\omega}_2)+e^{i \varphi}{\phi_\mathrm{f.d.}}({\omega}_2,{\omega}_1)].\label{EQ:FESpec}
\end{align}
with $\mathcal{N}^{-2}=2\mu(1+\cos(\varphi)\mathrm{sinc}(2\mu/\xi))$. In all expressions (\ref{EQ:GG}), (\ref{EQ:FE1}) and (\ref{EQ:FESpec}) $\xi$ is the single-photon bandwidth. Inserting (\ref{EQ:GG}), (\ref{EQ:FE1}) and (\ref{EQ:FESpec}) into (\ref{EQ:d}) yields 
\begin{subequations}
\label{EQ:IP}
\begin{align}
d_{\phi_{\mathrm{HOM}}}(\tau_1,\tau_2)=&e^{-\xi^2\Delta\tau^2}.\label{EQ:d1}
\\
d_{\phi_{\mathrm{f.d}}}(\tau_1,\tau_2)=&S_{\mu\xi}(\tau_1,\tau_2)\label{EQ:resFD}
\\
d_{\phi_{\mathrm{f.e}}}(\tau_1,\tau_2)=&2\mu\mathcal{N}^2[R_{\mu\xi}(\tau_1,\tau_2)+S_{\mu\xi}(\tau_1,\tau_2)]\label{EQ:HOM_FE}
\end{align}
\end{subequations}
with
\begin{subequations}
\begin{align}
S_{\mu\xi}(\tau_1,\tau_2)=&\frac{\sin\left(\frac{2\mu}{\xi}\mathrm{tri}\left(\frac{\xi\Delta\tau}{2}\right)\right)}{\frac{2\mu}{\xi}}\label{EQ:SF}
\\
R_{\mu\xi}^\varphi(\tau_1,\tau_2)=&\cos(\mu\Delta\tau-\varphi)\mathrm{tri}\left(\frac{\xi\Delta\tau}{2}\right).\label{EQ:RF}
\end{align}
\end{subequations}
where 
\begin{align}
\mathrm{tri}(x)=
\begin{cases}
1-|x| \ \ \ \ \ &\text{  if } |x|\leq 1 \\0 &\text{  if } |x|> 1
\end{cases}
\end{align}
is the \textit{triangular function}.

We want to briefly discuss some of the most important features of the interference patterns (\ref{EQ:IP}) of the three light sources. In the absence of spectral entanglement photon anti-bunching (i.e. $P^c > 1/2$) cannot occur \citep{HOMTheory}. In particular this holds true for the frequency uncorrelated photon source of the original HOM-experiment \citep{HOM} as seen from (\ref{EQ:d1}). This contrasts with the other two interference pattern, which both show a periodic change between photon bunching and photon anti-bunching in the delay $\Delta\tau$ with oscillation frequency $\mu$. This phenomenon is termed quantum beating \citep{ou1988observation} and is characteristic in the presence of spectral entanglement. Indeed both spectra (\ref{EQ:FE1}) and (\ref{EQ:FESpec}) contain spectral entanglement as they do not factorize into single-particle spectral \citep{ECKERT200288}. However, there is a striking difference between frequency detuned and frequency entangled photons, namely that the interference term (\ref{EQ:resFD}) of frequency detuned photons vanishes for increasing frequency separations $\mu$ where the one of frequency entangled photons (\ref{EQ:HOM_FE}) does not. As extensively discussed in \citep{Barzel2} the physical reason for this is the increasing spectral distinguishability of frequency detuned photons, which grows for larger values of $\mu$ and thus suppresses both photon bunching and photon anti-bunching. Mathematically perfect spectral distinguishability is expressed through a vanishing overlap of the spectral wave function with itself under exchange of the function arguments, i.e. $\lim_{\mu\rightarrow\infty}\int d\omega_1 d\omega_2 \phi_\mathrm{f.d.}(\omega_1,\omega_2)\phi_\mathrm{f.d.}^*(\omega_2,\omega_1) = 0$, which forces the interference term (\ref{EQ:d}) to vanish. This is different in the case of frequency entangled photons, which preserve their spectral indistinguishability, also for larger values of $\mu$.
Therefore, in contrast to the other two cases with frequency entangled photons we can achieve the following: in the limit $\mu\rightarrow\infty$, the occurrence of photon bunching or photon anti-bunching is highly sensitive to a change in the optical delay $\Delta\tau=\tau_1-\tau_2$. This fact will become important when we consider relativistic effects on HOM-interference in Section~\ref{sec:RELHOM}.

\section{\label{sec:REL} Relativistic deformation of spectral the wave function}
We now include relativistic effects. More concretely, we show that different observers would assign different spectral wave functions to the same (multi-) photon quantum state. A more detailed view on the topic in the single-particle sector is left to the literature \citep{birrell1984quantum,Bruschi1,Bruschi2}.

\subsection{Quantum fields in curved spacetime}
The starting point is to describe the light field as an uncharged massless real scalar field $\phi$, for the sake of simplicity and without loss of generality. The field $\phi$ obeys the massless Klein-Gordon equation
\begin{align}
(\sqrt{-g})^{-1}\partial_\mu(\sqrt{-g}g^{\mu\nu}\partial_\nu \phi)=0,
\end{align}
where $g_{\mu\nu}$ is the metric tensor of the underlying spacetime and $g$ is its determinant \cite{birrell1984quantum}.\footnote{We employ Einstein summation convention. The metric has signature $(-,+,+,+)$.} If the spacetime is stationary it possess a timelike Killing vector field which is $\partial_\xi$, and  elements of the metric tensor are independent of the time $\xi$. In our case we consider that the time parameter coincides with the coordinate time $t$, and therefore invariance of the metric under translations induced by the Killing vector field $\partial_t$ implies energy conservation. In turn, this allows one to decompose any solution of the Klein-Gordon equation into energy eigenstates $\phi_{\boldsymbol{k}}$ as $\phi=\int d^3 k[\alpha_{\textbf{k}}\phi_{\textbf{k}}+\alpha^*_{\textbf{k}}\phi_{\textbf{k}}^*]$. Here $\mathbf{k}$ collects the relevant quantum numbers, $\{\alpha_{\boldsymbol{k}}\}$ are Foruier coefficients for the classical field expansion, and $i\partial_t\phi_{\boldsymbol{k}}=\omega_{\boldsymbol{k}}\phi_{\boldsymbol{k}}$ is the eigenvector equation for the modes $\phi_{\boldsymbol{k}}$, where $\omega_{\boldsymbol{k}}$ is the (positive) eigenvalue. Canonical quantization promotes the expansion coefficients $\alpha^*_{\boldsymbol{k}}$ and $\alpha_{\boldsymbol{k}}$ of this decomposition into creation and annihilation operators $\hat{a}^\dag_{\boldsymbol{k}}$ and $\hat{a}_{\boldsymbol{k}}$ respectively. These operators satisfy the canonical commutation relations $[\hat{a}_{\boldsymbol{k}},\hat{a}^\dag_{\boldsymbol{k}'}]=\delta^3(\boldsymbol{k}-\boldsymbol{k}')$. 

\subsection{Single-photon wave packet deformation}
Full characterization of a realistic photon would require the use of electrodymanics in curved spacetime, which can be extremely cumbersome and might obfuscate the final results. Here we choose to follow another route. 

In the past decade a new body of work has developed the theory of photons propagating in (weakly) curved spacetime from different perspectives \citep{Bruschi1,Bruschi2,Exirifard:Culf:2021,Exirifard:Karimi:2021}. We assume that a photon can be indeed modelled by using a massless scalar field, and that it is effectively strongly confined to the direction of propagation in a (weakly curved) static spacetime. In this way, we can effectively assume that the photon is localized along a litghtlike path, and therefore we can ignore the deformation effects that occur in the perpendicular directions \cite{Exirifard:Culf:2021,Exirifard:Karimi:2021}. Nevertheless, we expect that the gravitational redshift will still affect the photon, and this effect in the context of a localized wavepacket has been pioneered in the literature \citep{Bruschi1,Bruschi2}.

We therefore characterize a single photon generated by a photon source through the spectral emission profile ${\Phi}_\mathrm{}({\omega}_\mathrm{})$ by the creation operator
\begin{align}
\hat{A}^\dagger(t_\mathrm{})=\mathcal{N}_{\psi_1}\int d{\omega}_\mathrm{}\,{\Phi}_\mathrm{}({\omega}_\mathrm{}) e^{i\omega t} \hat{a}^\dagger_{{\omega}},\label{EQ:aS}
\end{align}
where $t$ is the coordinate time and $\mathcal{N}_{\psi_1}$ is a normalization constant specified further below. The corresponding single-photon quantum state is
\begin{align}
\ket{\psi_1(t)}=\hat{A}^\dagger(t)\ket{0}.\label{EQ:PK0}
\end{align}
The same photon as seen from a moving observer $\mathrm{K}$ exploring the proper time $\tau_\mathrm{K}$ would be described by the creation operator $\hat{A}^\dagger_{\mathrm{K}}(\tau_\mathrm{K})$ which has been computed in the literature \citep{Bruschi1,Bruschi2}, and reads
\begin{align}
\hat{A}^\dagger_{\mathrm{K}}(\tau_\mathrm{K})=\int\displaylimits_0^\infty d{\omega}_\mathrm{K}{\Phi}^\mathrm{K}({\omega}_\mathrm{K})e^{i\omega_\mathrm{K}\tau_\mathrm{K}}\hat{a}^\dagger_{{\omega}_\mathrm{K}}.\label{EQ:EQ}
\end{align}
Thus different observers $\mathrm{K}$ will assign different spectral distributions ${\Phi}^\mathrm{K}({\omega}_\mathrm{K})$ to the very same photon. In particular, they will define the single photon quantum states $\ket{\psi_1(\tau_{\mathrm{K}})}$ as
\begin{align}
\ket{\psi_1(\tau_{\mathrm{K}})}=\hat{A}^\dagger_{\mathrm{K}}(\tau_\mathrm{K})\ket{0},\label{EQ:PK}
\end{align}
where the relation between the spectra ${\Phi}^\mathrm{K}({\omega}_\mathrm{K})$ of two observers Alice and Bob, with K$=$A,B respectively, has been computed in the literature \citep{Bruschi1}, and is given by
\begin{align}
{\phi}^\mathrm{B}({\omega}_\mathrm{B})=\sqrt{\frac{1}{1+z_{\mathrm{AB}}}}{\Phi}^\mathrm{A}\left({\omega_\mathrm{A}}{}\right).\label{EQ:B1}
\end{align}
Here we have used the defining relation between the frequencies as measured locally by Alice and Bob, which reads 
\begin{align}
\omega_\mathrm{A}=\frac{\omega_\mathrm{B}}{1+z_{\mathrm{AB}}},\label{EQ:B2}
\end{align}
where $z_\mathrm{AB}$ is the mutual redshift between observer $\mathrm{A}$ and $\mathrm{B}$ as defined in \citep{PhysRevD.95.104037}. The relations (\ref{EQ:B1}) and (\ref{EQ:B2}) follow from requiring Equation~(\ref{EQ:PK}) to yield proportional results for $\mathrm{K=A}$ and $\mathrm{K=B}$ (i.e. $\ket{\psi_1(\tau_{\mathrm{B}})}\propto\ket{\psi_1(\tau_{\mathrm{A}}(\tau_{\mathrm{B}}))}$) together with the normalization condition 
\begin{align}
\left(\mathcal{N}_{\psi_1}\right)^{-2}=\int d\omega_\mathrm{K}|{\Phi}^\mathrm{K}({\omega}_\mathrm{K})|^2
\end{align}
which ensures $\braket{\psi_1(\tau_\mathrm{K})|\psi_1(\tau_\mathrm{K})}=1$ in all frames.
Note that Equations~(\ref{EQ:EQ})-(\ref{EQ:B2}) are only valid approximations in a weakly curved spacetime \citep{Bruschi1,Bruschi2}. They become exact in the case that $\mathrm{K=A,B}$ are inertial observers in flat a spacetime \citep{carroll2019spacetime}.

As mentioned above, we here only consider the spectral aspects of the light field since we use the Klein-Gordon equation, which is a good approximation to the longitudinal (or transverse) modes of the electromagnetic field in scalar optics. A description of the observer dependent transformation properties of other DOFs of the light field, such as the helicity or the light's orbital momentum, would amount for a more elaborated treatment of the electromagnetic field. For this one would have to quantize Maxwell's equations on a curved spacetime background.

However, we proceed with a description within scalar optics, and (artificially) incorporate the other photonic DOFs by describing single photon states using a generalized version of the creation operator (\ref{EQ:EQ}), which reads
\begin{align}
\hat{A}^\dagger_{\mathrm{K}}(\tau_\mathrm{K})=\int\displaylimits_0^\infty d{\bm{\omega}}_\mathrm{K}{\Phi}^\mathrm{K}({\bm{\omega}}_\mathrm{K})e^{i\omega_\mathrm{K}\tau_\mathrm{K}}\hat{a}^\dagger_{{\bm{\omega}}_\mathrm{K}},\label{EQ:EQsigma}
\end{align}
where we defined the entire set of photonic DOFs as $\bm{\omega}_\mathrm{K}:=\{\omega_\mathrm{K},\bm{\sigma}\}$, and $\bm{\sigma}$ is the set of photonic DOFs excluding for the photons frequency $\omega_\mathrm{K}$. Clearly, these DOFs are those potentially measured by from an observer K. The observer dependent photonic wave function is ${\Phi}^\mathrm{K}({\bm{\omega}}_\mathrm{K})$, which we alternatively write as ${\Phi}^\mathrm{K}_{\bm{\sigma}}({\omega}_\mathrm{K})={\Phi}^\mathrm{K}({\bm{\omega}}_\mathrm{K})$.

\subsection{Two-photon wave packet deformation}
We now turn to the description of two photons. In analogy to (\ref{EQ:PK0}) we can write a general two-photon state, which is a solution of the Klein-Gordon equation (with a stationary metric) as
{\small
\begin{align}
\ket{\psi_2(t)}=\mathcal{N}_{\psi_2}\int & {d{\bm{\omega}_1} d{\bm{\omega}_2}}\ {\Phi}({\bm{\omega}_1},{\bm{\omega}_2})
e^{i\omega_1t}e^{i\omega_\mathrm{2}t}|1_{{\bm{\omega}}_\mathrm{1}}1_{{\bm{\omega}}_\mathrm{2}}\rangle,
\label{EQ:JPT}
\end{align}
}
where $|1_{{\bm{\omega}}_\mathrm{1}}1_{{\bm{\omega}}_\mathrm{2}}\rangle:= \hat{a}^\dagger_{{\bm{\omega}}_\mathrm{1}}\hat{a}^\dagger_{{\bm{\omega}}_\mathrm{2}}\ket{0}$ and $\mathcal{N}_{\psi_2}$ being a normalization constant, which we specify further below. 

Now we work out how this quantum state is characterized by two observers $\mathrm{K}_1$ and $\mathrm{K}_2$ both moving in a curved spacetime. Our aim is to find a generalization of the transformation (\ref{EQ:B1}) and (\ref{EQ:B2}) for the two-photon spectral wave function $\Phi^{\mathrm{K_1}\mathrm{K_2}}({\bm{\omega}}_{\mathrm{K_1}},{\bm{\omega}}_{\mathrm{K_2}})$ as seen by the observer pair $\mathrm{K_1,K_2}$. However, a straightforward generalization to the single particle case here is not possible as quantum fields (and in particular those of several distant particles) in quantum field theory are inherently non-local objects where the notion of an observer from general relativity is strictly local \citep{Bruschi1}.

In a first attempt we follow a simplified approach and consider inertial observers moving in a flat spacetime geometry, which is the extreme case of considering a weakly curved spacetime and small accelerations. This has the advantage that all observers in this scenario share a common vacuum state and therefore a common notion of particles, and thus circumvents the conceptional problems related to non-locality in curved spacetimes. Nevertheless, also in this simplified approach the mutual redshift between all considered observers among each other and with respect to the photon source enters in a nontrivial way, such that one can study the most basic spectral aspects of HOM-interferometry under relativistic influences. Once we have worked out how the measurement outcome of a HOM-experiment is influenced by a mutual redshift between the observers among each other and with respect to the photon source, we can replace the redshift in the resulting formulas by the gravitational redshift to get an estimate of general relativistic influences on the measurement outcome of HOM-experiments under the assumption of weakly curved spacetimes and small accelerations. It is crucial to note that from now on, since we consider two photons, all measurements have to be considered with respect to two observers, or an \textit{observer pair}. This also includes the special case where the two considered observers in a pair coincide.

In analogy to (\ref{EQ:EQsigma}) we make the ansatz to characterize two-photon states as seen from two observers $\mathrm{K_1}$ and $\mathrm{K_2}$ (the corresponding observer pair is denoted as $\mathrm{K_1},\mathrm{K_2}$) as
\begin{align}
\ket{\psi_2(\tau_{\mathrm{K}_1},\tau_{\mathrm{K}_2})}=&\mathcal{N}_{\psi_2}\int\displaylimits_0^\infty  d{\bm{\omega}}_\mathrm{K_1} d{\bm{\omega}}_\mathrm{K_2}\ {\Phi}^{\mathrm{K_1}\mathrm{K_2}}({\bm{\omega}}_{\mathrm{K_1}},{\bm{\omega}}_{\mathrm{K_2}})\nonumber\\
&\times e^{i\omega_\mathrm{K_1}\tau_\mathrm{K_1}}e^{i\omega_\mathrm{K_2}\tau_\mathrm{K_2}} a^\dagger_{{\bm{\omega}}_\mathrm{K_1}}a^\dagger_{{\bm{\omega}}_\mathrm{K_2}}\ket{0},
\label{EQ:JP}
\end{align}
where now ${\Phi}^\mathrm{K_1K_2}({\bm{\omega}}_{\mathrm{K_1}},{\bm{\omega}}_{\mathrm{K_2}})\equiv{\Phi}_{\bm{\sigma}_1\bm{\sigma}_2}^\mathrm{K_1K_2}({{\omega}}_{\mathrm{K_1}},{{\omega}}_{\mathrm{K_2}})={\Phi}^\mathrm{K_1K_2}({\bm{\omega}}_{\mathrm{K_1}},{\bm{\omega}}_{\mathrm{K_2}})$ is the joint two-photon wave function as seen from the observers $\mathrm{K}_1$ and $\mathrm{K}_2$, and $\mathcal{N}_{\psi_2}$ is a normalization constant in order to fulfill $\braket{\psi_2|\psi_2}=1$ in all frames, which is determined by
\begin{align}
\left(\mathcal{N}^2_{\psi_2}\right)^{-2}=\int& d\bm{\omega}_{\mathrm{K}_1}d\bm{\omega}_{\mathrm{K}_2} |{{\Phi}^\mathrm{K_1K_2}_\mathrm{S}}({\bm{\omega}}_{\mathrm{K_1}},{\bm{\omega}}_{\mathrm{K_2}})|^2,
\label{EQ:normSimple}
\end{align}
where we defined the symmetrized two-photon wave function ${{\Phi}^\mathrm{K_1K_2}_\mathrm{S}}({\bm{\omega}}_{\mathrm{K_1}},{\bm{\omega}}_{\mathrm{K_2}})=({{\Phi}^\mathrm{K_1K_2}}({\bm{\omega}}_{\mathrm{K_1}},{\bm{\omega}}_{\mathrm{K_2}})+{{\Phi}^\mathrm{K_1K_2}}({\bm{\omega}}_{\mathrm{K_2}},{\bm{\omega}}_{\mathrm{K_1}}))/\sqrt{2}$.
The normalization constant (\ref{EQ:normSimple}) reflects the fact that only the symmetric part of the photonic wave function is of physical relevance as photons are bosons. Indeed one can use the canonical commutator relations to rewrite the two-photon state (\ref{EQ:JP}) solely in terms of the symmetrized wave function ${{\Phi}^\mathrm{K_1K_2}_\mathrm{S}}({\bm{\omega}}_{\mathrm{K_1}},{\bm{\omega}}_{\mathrm{K_2}})$. Also if one does not use symmetrized photonic wave functions the second quantization formalism automatically accounts only for the symmetric part of the photonic wave function. 

In Equation~(\ref{EQ:JP}), the parameters $\tau_{\mathrm{K_1}}$ and $\tau_{\mathrm{K_2}}$ are the lapse that the respective observers associate to the separate phase in their local time. The ansatz (\ref{EQ:JP}) is exact in case of inertial observers $\mathrm{K}_1$ and $\mathrm{K}_2$ in a flat spacetime \citep{carroll2019spacetime}. Now we want to analyze how the joint spectral profile of two photons as seen from a given pair of inertial observers, say $\mathrm{K_1=A_1}$ and $\mathrm{K_2=A_2}$, changes when seen from a different pair of inertial observers, say $\mathrm{K_1=B_1}$ and $\mathrm{K_2=B_2}$. In analogy to the single photon case we require 
\begin{align}
\ket{\psi_2(\tau_{\mathrm{B}_1},\tau_{\mathrm{B}_2})}\propto \ket{\psi_2(\tau_{\mathrm{A}_1}(\tau_{\mathrm{B}_1}),\tau_{\mathrm{A}_2}(\tau_{\mathrm{B}_2}))}\label{EQ:prop}
\end{align}
under the normalization condition (\ref{EQ:normSimple}). This yields the relation
\begin{align}
{\Phi}^\mathrm{B_1B_2}_{\bm{\sigma}_1\bm{\sigma}_2}({{\omega}}_\mathrm{B_1},{\omega}_\mathrm{B_2})=&\sqrt{\frac{1}{(1+z_{\mathrm{A_1B_1}})(1+z_{\mathrm{A_2B_2}})}}
\nonumber\\
&\times
{\Phi}^\mathrm{A_1A_2}_{\bm{\sigma}_1\bm{\sigma}_2}({{\omega}}_\mathrm{A_1},{\omega}_\mathrm{A_2}),\label{EQ:R1}
\end{align}
where we have 
\begin{align}
{\omega}_{\mathrm{A}_i}=\frac{{{\omega}}_{\mathrm{B}_j}}{1+z_{\mathrm{A}_i\mathrm{B}_j}},\ \ \ i,j=1,2\label{EQ:R2}
\end{align}
and $z_{\mathrm{A}_i\mathrm{B}_j}$ as before being the mutual redshift between the two observers $\mathrm{A}_i$ and $\mathrm{B}_j$. One can check that photonic wave functions obeying the transformation behavior of Equations~(\ref{EQ:R1}) and (\ref{EQ:R2}) fulfill the requirements defined by (\ref{EQ:normSimple}) and (\ref{EQ:prop}). 

Note that Equation~(\ref{EQ:R1}) only depends on the mutual redshifts $z_{\mathrm{A}_1\mathrm{B}_1}$ and $z_{\mathrm{A}_2\mathrm{B}_2}$ of the first and second observer in an observer pair respectively but not on the redshifts $z_{\mathrm{A}_1\mathrm{B}_2}$ and $z_{\mathrm{A}_2\mathrm{B}_1}$. It might therefore appear that our formalism is able to extract the which-path information, and that our results may vary if one exchanges the two observers of an observer pair. However, this is not the case. To see this we first note that the prefactor of (\ref{EQ:R1}) is invariant under the exchange of the observers of one observer pair, because $(1+z_{\mathrm{A_1B_1}})(1+z_{\mathrm{A_2B_2}})=(1+z_{\mathrm{A_1B_2}})(1+z_{\mathrm{A_2B_1}})=(1+z_{\mathrm{A_1B_1}})(1+z_{\mathrm{A_2B_1}})$, which follows from the transitivity property of the redshift \citep{wald2010general} and can be inferred from (\ref{EQ:R2}). Secondly, as only the symmetrized (transformed) photonic wave function $({\Phi}^\mathrm{B_1B_2}_{\bm{\sigma}_1\bm{\sigma}_2}({{\omega}}_\mathrm{B_1} {\omega}_\mathrm{B_2})+{\Phi}^\mathrm{B_1B_2}_{\bm{\sigma}_2\bm{\sigma}_1}({{\omega}}_\mathrm{B_2},{\omega}_\mathrm{B_1}))\sqrt{2}$ enters the calculations, it follows that the expression (\ref{EQ:R1}) is fully invariant under the exchange of $\mathrm{A_1}$ with $\mathrm{A_2}$ and $\mathrm{B_1}$ with $\mathrm{B_2}$. 

The relations (\ref{EQ:R1}) and (\ref{EQ:R2}) quantify the deformation of the joint spectral distribution of the two-photon state (\ref{EQ:JP}), when seen from different pairs of inertial observers in a flat spacetime. Note that the here presented framework is also capable to describe the spectral aspects of moving interference experiments like the Sagnac-interferometer \citep{restuccia2019photon,Sagnac}. Also simple phenomena like the change in frequency of a light beam when it is reflected on a moving mirror can be described with this framework. Moreover, the generalization to three, four and more photons is straightforward.

\begin{figure}[t]\centering
  \includegraphics[width=0.99\linewidth]{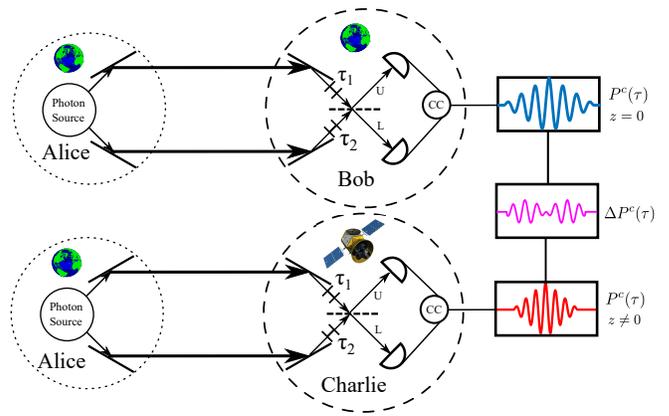}
  \caption{Top: HOM-experiment without relativistic influence (i.e. vanishing redshift $z=0$). Bottom: HOM-experiment with relativistic influence (i.e. non-vanisching redshift $z\neq 0$). The dotted circles indicate the location, where the photons are generated (in both cases by Alice on Earth). The location where the coincidence statistics are recorded (by Bob on Earth in the upper plot and on a moving satellite by Charlie in the lower plot) are indicated by dashed circles, which we term detection instance. The graphs on the right side represent the coincidence interference pattern, which is recorded at the respective detection instance Bob and Charlie. Bob on Earth records the blue curve and Charlie on a satellite records the red one. Both Bob and Charlie apply optical delays $\tau_1$ and $\tau_2$ in units of their local proper times respectively. Due to the different redshifts between the photon source and the respective detection instance slightly different coincidence interference patterns are recorded and their difference is displayed by the purple curve.}
\label{fig:RelHOM}
\end{figure}

\section{\label{sec:RELHOM} Relativistic Influence on HOM-Interference}
We are now equipped with the tools necessary to quantify the changes of the interference pattern of a HOM-experiment subject to the relativistic frequency shift. The validity of Equation~(\ref{EQ:P}) in a local inertial frame is well confirmed by many experiments \citep{HOM,
Zeillinger,
Zeillinger2,
liu2020sub,
ou1988observation}. Thus we assume that Equation~(\ref{EQ:P}) locally holds true, that is, we replace the optical delays $\tau_1$ and $\tau_2$ in (\ref{EQ:P}) by optical delays $\tau_{\mathrm{K}_1}$ and $\tau_{\mathrm{K}_2}$ as measured in the proper time of the respective observer $\mathrm{K}_1$ and $\mathrm{K}_2$, applying the respective optical delay. Furthermore, we replace the two-photon wave function $\Phi_{\bm{\sigma}_1\bm{\sigma}_2}(\omega_1,\omega_2)$, which enters Equation~(\ref{EQ:P}) by the corresponding two-photon wave function $\Phi_{\bm{\sigma}_1\bm{\sigma}_2}^{\mathrm{K}_1\mathrm{K}_2}(\omega_1,\omega_2)$ as seen by the observer pair $\mathrm{K}_1,\mathrm{K}_2$. Accordingly we can write the relativistic version of (\ref{EQ:P}) as
\begin{align}
\tilde{P}^{\mathrm{K}_1\mathrm{K}_2}_{\bm{\sigma}_1\bm{\sigma}_2}(\tau_{\mathrm{K}_1},\tau_{\mathrm{K}_2})=&\int d \omega_1d \omega_2\ \left[|\Phi^{\mathrm{K}_1\mathrm{K}_2}_{\bm{\sigma}_1\bm{\sigma}_2}(\omega_1,\omega_2)|^2\right.
\nonumber
\\
&\hspace*{20mm} +\left.|\Phi^{\mathrm{K}_1\mathrm{K}_2}_{\bm{\sigma}_2\bm{\sigma}_1}(\omega_2,\omega_1)|^2\right]
\nonumber
\\
&\hspace*{-20mm}+2\Re\left\{\int d \omega_1d \omega_2 \Phi^{\mathrm{K}_1\mathrm{K}_2}_{\bm{\sigma}_1\bm{\sigma}_2}(\omega_1,\omega_2){\Phi^{\mathrm{K}_1\mathrm{K}_2}_{\bm{\sigma}_2\bm{\sigma}_1}}^*(\omega_2,\omega_1)\right.
\nonumber
\\
&\times\left. e^{-i(\omega_1-\omega_2)(\tau_{\mathrm{K}_1}-\tau_{\mathrm{K}_2})}\right\}.\label{EQ:PRel}
\end{align}
To obtain the normalized probabilities of a certain detection event, which is characterized by $\bm{\sigma}_1$ and $\bm{\sigma}_2$ one has to compute
\begin{align}
P^{\mathrm{K}_1\mathrm{K}_2}_{\bm{\sigma}_1\bm{\sigma}_2}(\tau_{\mathrm{K}_1},\tau_{\mathrm{K}_2})=\frac{\tilde{P}^{\mathrm{K}_1\mathrm{K}_2}_{\bm{\sigma}_1\bm{\sigma}_2}(\tau_{\mathrm{K}_1},\tau_{\mathrm{K}_2})}{\int d\bm{\sigma}_1d\bm{\sigma}_2\, \tilde{P}^{\mathrm{K}_1\mathrm{K}_2}_{\bm{\sigma}_1\bm{\sigma}_2}(\tau_{\mathrm{K}_1},\tau_{\mathrm{K}_2})}.\label{EQ:normPRel}
\end{align}
Note that Equation~(\ref{EQ:PRel}) is invariant under the exchange of $\tau_{\mathrm{K_1}}$ and $\tau_{\mathrm{K_2}}$. This means that our formalism does not depend on the information of which observer of an observer pair applies which optical delay.

Equations (\ref{EQ:PRel}) and (\ref{EQ:normPRel}) are central results of the present paper as they capture the observer dependence of the spectral aspects of HOM-interference. The dependence of the proper times $\tau_{\mathrm{K}_1}$ and $\tau_{\mathrm{K}_2}$ describes how the interference pattern changes when recorded by different observer pairs and the dependence on the wave function $\Phi^{\mathrm{K}_1\mathrm{K}_2}_{\bm{\sigma}_1\bm{\sigma}_2}(\omega_1,\omega_2)$ as seen by the respective observer pair reflects the influence of a redshift between the light source and the observers on the interference pattern.

We employ Equations (\ref{EQ:R1}),(\ref{EQ:PRel}) and (\ref{EQ:normPRel}) to obtain the relation between the interference patterns obtained by two distinct observer pairs
\begin{align}
P^{\mathrm{B_1B_2}}_{\bm{\sigma}_1\bm{\sigma}_2}(\tau_\mathrm{B_1},\tau_\mathrm{B_2})=P^{\mathrm{A_1A_2}}_{\bm{\sigma}_1\bm{\sigma}_2}(\tau_\mathrm{A_1},\tau_\mathrm{A_2}),\label{EQ:relAB}
\end{align}
where the relation between the delays $\tau_{\mathrm{A}_i}$ and $\tau_{\mathrm{B}_i}$ reads 
\begin{align}
\tau_{\mathrm{A}_i}=(1+z_{\mathrm{A}_i\mathrm{B}_i})\tau_{\mathrm{B}_i},\ i=1,2.\label{EQ:tAB}
\end{align}
Note that Equation~(\ref{EQ:tAB}) is precisely the relation that tells how much the (proper) time $\tau_{\mathrm{A}_i}$ has evolved on the world line of observer $\mathrm{A}_i$, when a (proper) time of $\tau_{\mathrm{B}_i}$ has evolved on the world line of observer $\mathrm{B}_i$ \citep{philipp2017definition}. Equations~(\ref{EQ:relAB}) and (\ref{EQ:tAB}) relate the interference patterns of two HOM-experiments, which are done by two different observer pairs $\mathrm{A}_1,\mathrm{A}_2$ and $\mathrm{B}_1,\mathrm{B}_2$ featuring a mutual redshift to each other.

In the following we always assume that Alice operates the two-photon source and distributes the photons. However, also Alice can measure the HOM-interference pattern of the photon source. If the photon source, all reflecting elements and detectors rest in the local inertial frame of Alice, she defines an observer pair $\mathrm{A_1}$, $\mathrm{A_2}$. Therefore $P^{\mathrm{A_1A_2}}_{\bm{\sigma}_1\bm{\sigma}_2}(\tau_\mathrm{A_1},\tau_\mathrm{A_2})$ is the corresponding interference pattern of the photon source in the local rest frame of the same, i.e., the source specific interference pattern of the considered photon source without relativistic effects. However, Alice can also send the photons to another observer pair $\mathrm{B_1}$, $\mathrm{B_2}$ (Bob), who in this case would record the interference pattern $P^{\mathrm{B_1B_2}}_{\bm{\sigma}_1\bm{\sigma}_2}(\tau_\mathrm{B_1},\tau_\mathrm{B_2})$, which is related to interference pattern $P^{\mathrm{A_1A_2}}_{\bm{\sigma}_1\bm{\sigma}_2}(\tau_\mathrm{A_1},\tau_\mathrm{A_2})$ via Equations (\ref{EQ:relAB}) and (\ref{EQ:tAB}). Alice might provide the photons to yet another observer pair $\mathrm{C_1}$, $\mathrm{C_2}$ (Charlie),  who would record the interference pattern $P^{\mathrm{C_1C_2}}_{\bm{\sigma}_1\bm{\sigma}_2}(\tau_\mathrm{C_1},\tau_\mathrm{C_2})$, which is related to interference pattern $P^{\mathrm{A_1A_2}}_{\bm{\sigma}_1\bm{\sigma}_2}(\tau_\mathrm{A_1},\tau_\mathrm{A_2})$ by replacing $\mathrm{B_1}$ and $\mathrm{B_2}$ in Equations (\ref{EQ:relAB}) and (\ref{EQ:tAB}) by $\mathrm{C_1}$ and $\mathrm{C_2}$. From the transitivity property of the redshift (which we discussed below Equation (\ref{EQ:R2})) we can obtain the relations between the interference pattern recorded by Bob's and Charlie's observer pairs, and they are given by simply replacing $\mathrm{A_1}$ and $\mathrm{A_2}$ in Equations (\ref{EQ:relAB}) and (\ref{EQ:tAB}) by $\mathrm{C_1}$ and $\mathrm{C_2}$. Interestingly this implies that the relation between the interference patterns respectively recorded by Bob's and Charlie's observer pair does not depend on the mutual redshift w.r.t. the photon source (i.e. to Alice' observer pair) but only on the mutual redshift between Bob's and Charlies's observer pairs. This might be interesting for Geodesy, where Bob and Charlie are provided with photons from a common light source and can infer information about the geometry of spacetime, solely by comparing their HOM-interference patterns with each other, without knowledge about the actual state of motion and the position of the light source. However, each of the interference pattern recorded by Bob and Charlie depend on the mutual redshift with respect to the photon source, which can be quantified by considering the observer pairs $\mathrm{B}_1,\mathrm{B}_2$ or $\mathrm{C}_1,\mathrm{C}_2$ on one side and the pair $\mathrm{A}_1,\mathrm{A}_2$ on the other side of Equations~(\ref{EQ:relAB}) and (\ref{EQ:tAB}).

 \begin{widetext}
\begin{figure*}[t]\centering
  \includegraphics[width=0.485\linewidth]{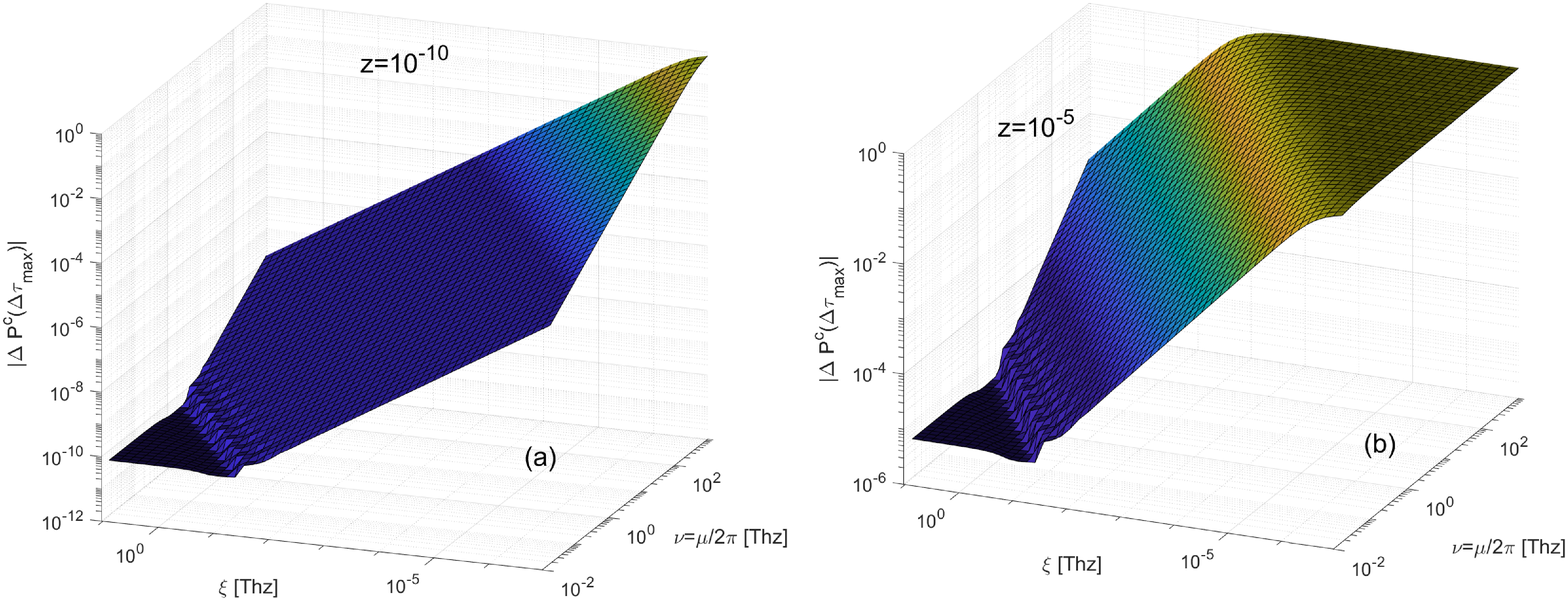}
  \includegraphics[width=0.485\linewidth]{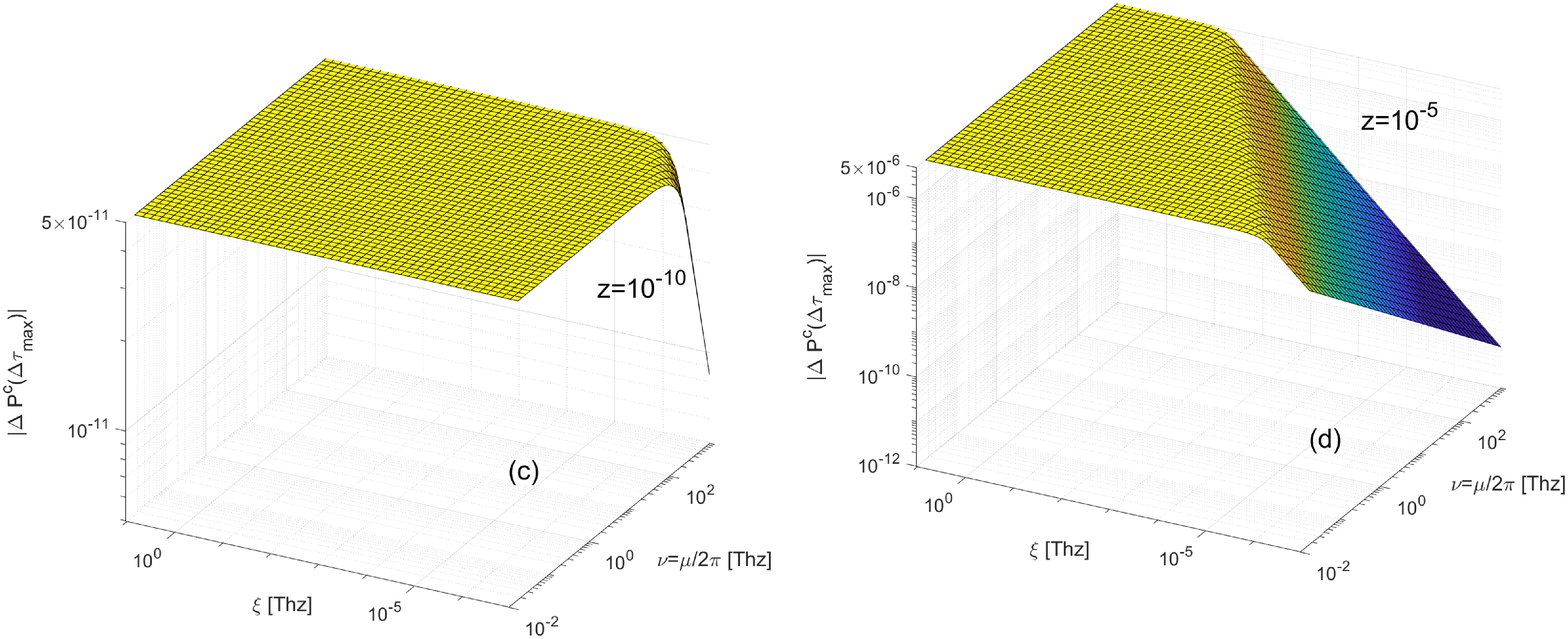}
  \caption{All plots show the absolute value of the maximum Discrepancy $|P^c(\Delta\tau_{\max})|$ (i.e. the impact of the relativistic redshift on HOM-interference) as a function of the frequency separation $\nu$ and the single photon band width $\xi$. The upper plots show the influence of a redshift of (a) $z=10^{-10}$ and (b) $z=10^{-5}$ on a HOM-experiment employing frequency entangled photons. The lower plots show the influence of a redshift of (c) $z=10^{-10}$ and (d) $z=10^{-5}$ on a HOM-experiment employing frequency-detuned polarization-entangled photons. It can be seen that in the case of frequency entangled photons adjusting the experimental parameters $\nu$ to higher values and $\xi$ to lower values amplifies the relativistic influence on the HOM-interference pattern. In case of frequency-detuned polarization-entangled photons the maximum discrepancy of $|P^c(\Delta\tau_{\max})|\approx z/2$ cannot be exceeded by a parameter adjustment. The plots (a) and (b) are computed with Equation~(\ref{EQ:HOM_FE}) for $\varphi=0$ and the plots (c) and (d) are computed with Equation~(\ref{EQ:resFD}).}
\label{fig:Z}
\end{figure*}
\end{widetext}

Note that Equation~(\ref{EQ:relAB}) also accounts for the situation in which the two photons explore different redshifts (i.e. $z_{\mathrm{A}_1\mathrm{B}_1}\neq z_{\mathrm{A}_2\mathrm{B}_2}$) when propagating between the photon source and the two different receivers. Such scenario would be of interest for the study of the spectral aspects of the Sagnac-effect in HOM-interference \citep{restuccia2019photon}, where the emission direction of the two photons from the photon source can have different angles with respect to the velocity vector of the photon source, thereby leading to different Doppler-shifts between the photons. 

For the rest of this work we consider a simplified situation, in which the two photons are received by the receiving observers with the same total redshift with respect to the emitter. This would occur, for instance, in the case when the two photons are generated by a source which rests on Earth and are sent to a moving satellite where they are received from a similar direction (to avoid additional Dopper shifts between the photons). Onboard the satellite, optical delays are employed (in units of the proper time of the satellite). This situation is illustrated in Figure~\ref{fig:RelHOM}. There it is seen that we consider a photon source held by Alice who is situated on Earth, and who sends photon pairs to Bob, who is situated on Earth as well. The mutual redshift between Alice and Bob is zero, i.e. $z_{\mathrm{A}_1\mathrm{B}_1} = z_{\mathrm{A}_2\mathrm{B}_2}=0$. In other words Bob is located in the same local rest frame as the photon source. In a second scenario, Alice sends the photon pair to Charlie, who is situated on a satellite and has a non-vanishing redshift with respect to Alice, i.e. $z_{\mathrm{A}_1\mathrm{C}_1} = z_{\mathrm{A}_2\mathrm{C}_2}=:z$. Both Bob and Charlie make a coincidence measurement and record the interference profile by tuning their optical delays (in units of their own proper times). The effects of the redshift between the photon source (dotted circle) and Charlie's detection instance (dashed circle) will lead to a difference (purple curve in Figure~\ref{fig:RelHOM}) of his interference pattern (red curve) with respect to the one recorded by Bob (blue curve). We now quantify this difference.

We are interested in the difference $\Delta P_{\bm{\sigma}_1\bm{\sigma}_2}(\tau_1,\tau_2)$ between the interference patterns of Bob and Charlie defined by
\begin{align}
\Delta P_{\bm{\sigma}_1\bm{\sigma}_2}(\tau_1,\tau_2)=P^{\mathrm{C_1C_2}}_{\bm{\sigma}_1\bm{\sigma}_2}(\tau_1,\tau_2)-P^{\mathrm{B_1B_2}}_{\bm{\sigma}_1\bm{\sigma}_2}(\tau_1,\tau_2).\label{EQ:DP}
\end{align} 
Notice that, as we have mentioned before, this difference is obtained operationally by recording the interference pattern \textit{with respect to the proper time}, and then exchanging the pattern to be compared.

Bob is located at the same height as Alice in the gravitational potential, and therefore has a vanishing redshift with respect to the photon source, i.e., we have $P^{\mathrm{B_1B_2}}_{\bm{\sigma}_1\bm{\sigma}_2}(\tau_1,\tau_2)=P^{\mathrm{A_1A_2}}_{\bm{\sigma}_1\bm{\sigma}_2}(\tau_1,\tau_2)$. Assuming a common mutual redshift between the input ports of Charlie's measurement apparatus with respect to the photon source, i.e., $z_{\mathrm{A}_1\mathrm{C}_1} = z_{\mathrm{A}_2\mathrm{C}_2}=z$, we know that $P^{\mathrm{C_1C_2}}_{\bm{\sigma}_1\bm{\sigma}_2}(\tau_1,\tau_2)=P^{\mathrm{A_1A_2}}_{\bm{\sigma}_1\bm{\sigma}_2}((1+z)\tau_1,(1+z)\tau_2)$, and Equation~(\ref{EQ:DP}) simplifies to 
\begin{align}
\Delta P_{\bm{\sigma}_1\bm{\sigma}_2}(\tau_1,\tau_2)=&P^{\mathrm{A_1A_2}}_{\bm{\sigma}_1\bm{\sigma}_2}((1+z)\tau_1,(1+z)\tau_2)\nonumber\\
&-P^{\mathrm{A_1A_2}}_{\bm{\sigma}_1\bm{\sigma}_2}(\tau_1,\tau_2).\label{EQ:DP2}
\end{align} 
It is clear from \eqref{EQ:DP2} that Bob's results act as a reference and therefore $\Delta P_{\bm{\sigma}_1\bm{\sigma}_2}(\tau_1,\tau_2)=0$ when $z=0$.

We then let $P^c(\tau_1,\tau_2)$ be the coincidence detection probability of the two-photon source in the local inertial frame of the photon source. From Equation~(\ref{EQ:DP2}), we can compute the difference $\Delta P^c(\tau_1,\tau_2)$ between the coincidence probabilities of two HOM-experiments with and without relativistic influence, which reads
{\small
\begin{align}
\Delta P^c(\tau_1,\tau_2)=P^c((1+z)\tau_1,(1+z)\tau_2)-P^c(\tau_1,\tau_2).
\end{align}
}
To see this, note that the coincidence probability $ P^c(\tau_1,\tau_2)$ is just a linear combination of the probabilities $P_{\bm{\sigma}_1\bm{\sigma}_2}(\tau_1,\tau_2)$ of specific detection events (with respect to any observer pair).

In case of all three light source types, which were considered in Section \ref{sec:HOM}, the coincidence probabilities were always functions of the delay $\Delta\tau=\tau_1-\tau_2$ and they were all of the form
\begin{align}
P^c(\Delta\tau)=\frac{1}{2}(1-d(\Delta\tau)),
\end{align}
where $d(\Delta\tau)$ depends on the joint spectral distribution of the respectively employed light source (c.f. Equations~(\ref{EQ:d})).
Then, the difference between the coincidence detection probabilities as a function of the redshift and the delay reads
\begin{align}
\Delta P^c(\Delta\tau)=\frac{1}{2}(d((1+z)\Delta\tau)-d(\Delta\tau)).\label{EQ:Pd}
\end{align}
In case of the traditional HOM-experiment employing polarization entangled photons with the joint spectral distribution (\ref{EQ:GG}) the difference of the coincidence detection probability reads
\begin{align}
\Delta P^c(\Delta\tau)=\frac{1}{2}(e^{-\xi^2(1+z)^2\Delta\tau^2}-e^{-\xi^2\Delta\tau^2}),\label{EQ:discraP}
\end{align}
which reaches its maximum at a delay of
\begin{align}
\Delta\tau_{\max}=\frac{1}{\xi}\sqrt{2\frac{\ln(1+z)}{z(2+z)}}.
\end{align}
The maximum value of the difference is independent of the bandwidth $\xi$ of the photon source, since there is only one time-scale $1/\xi$ that can be absorbed in the definition of time, and it reads
{\small
\begin{align}
\Delta P^c(\Delta\tau_{\max}) = \frac{1}{2}\left[ (1+z)^{-\frac{2(1+z)^2}{z(2+z)}}-(1+z)^{-\frac{2}{z(2+z)}}\right].\label{EQ:Pmax1}
\end{align}
}
This result shows that it would be extremely difficult to measure the influence of the redshift using the scheme described here in a practical scenario. In fact, a discrepancy of only $1\%$ (i.e. $\Delta P^c(\Delta\tau_{\max})=0.01$) would require the experimenter to achieve a redshift of the order of $z\approx 0.01$, which is about three orders of magnitude higher than the strongest redshifts (mainly caused by the Doppler-effect) between two satellites counter orbiting around the Earth ($z\approx10^{-5}$). Since Equation~(\ref{EQ:Pmax1}) is independent of the bandwidth $\xi$, achieving lower bandwidths will not amplify the influence of relativistic effects.

\begin{figure}[t]\centering
 \hspace*{-5mm} \includegraphics[width=1.1\linewidth]{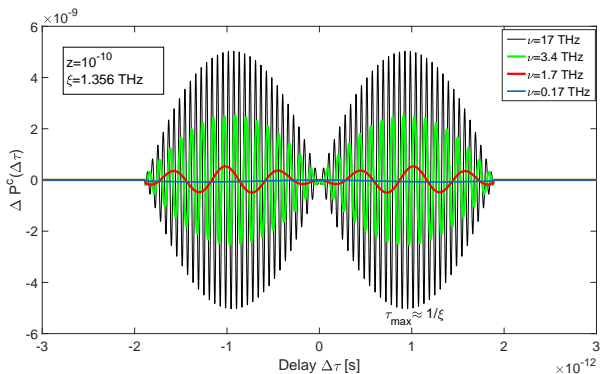}
  \caption{Difference $\Delta P^c(\Delta\tau)$ (see. Equation~(\ref{EQ:Pd})) between the coincidence detection probabilities of two HOM-experiments with frequency entangled photons (see. Equation~(\ref{EQ:FESpec}) for spectral profile) under the influence of a redshift $z=10^{-10}$ for different frequency separations $\nu$. The plot is shown for $\varphi=\pi$.}
\label{fig:pattern}
\end{figure}

This is however different if one employs a source of frequency entangled photons with a joint frequency profile of (\ref{EQ:FESpec}), which leads to the difference of the coincidence probability (\ref{EQ:Pd}), where one has to use the function $d(\Delta\tau)$ from Equation~(\ref{EQ:HOM_FE}). Figure~\ref{fig:pattern} shows the difference between the coincidence detection probabilities of two HOM-experiments with and without the influence of a redshift of $z=10^{-10}$ for different frequency separations $\nu=\mu/2\pi$ as a function of the delay $\Delta\tau=\tau_1-\tau_2$. It can be seen that the maximum discrepancy is achieved for a delay
\begin{align}
\Delta\tau_{\max}\approx\frac{1}{\xi}.\label{EQ:tmax}
\end{align}
The reason therefore is explained below. Furthermore it can be seen that the absolute value of the maximum discrepancy $|\Delta P^c(\Delta\tau_{\max})|$ increases with increasing frequency separation $\nu=\mu/2\pi$ (also explained below).

Figures~\ref{fig:Z} (a) and (b) show the result of the absolute value of the maximum (maximized over the delay $\Delta\tau$) difference $|\Delta P^c(\Delta\tau_{\max})|$ as a function of the experimental parameters $\xi$ and $\mu$ under the influence of a redshift $z=10^{-10}$ (Figure~\ref{fig:Z}(a)) and $z=10^{-5}$ (Figure~\ref{fig:Z}(b)). It can be seen that the influence of the relativistic redshift on the coincidence probability increases for higher frequency separations $\mu$ and for lower single photon bandwidths $\xi$ and for larger redshifts $z$. Therefore the most interesting parameter regime is the limit ${\mu/\xi\rightarrow\infty}$, where the function (\ref{EQ:HOM_FE}) is well approximated by $d(\Delta\tau)\approx R_{\mu\xi}^\varphi(\Delta\tau)$ with $R_{\mu\xi}^\varphi(\Delta\tau)$ taken from (\ref{EQ:RF}).
Thus the difference of the joint detection probabilities is approximately
\begin{align}
\Delta P^c(\Delta\tau)\approx\frac{1}{2}(R_{\mu\xi}^{\varphi}((1+z)\Delta\tau)-R_{\mu\xi}^{\varphi}(\Delta\tau)).\label{EQ:discApp}
\end{align}

Our numeric calculations yield that there is no qualitative difference between the results shown in Figure~\ref{fig:Z} for different values of $\varphi$ in the parameter region of interest $\mu\gg\xi$. This is why from here on we proceed our discussion for $\varphi=0$.
 
We see from (\ref{EQ:discApp}) for $\varphi=0$ that the (absolute value of the) difference of the cosine functions $\cos(\mu(1+z)\Delta\tau)-\cos(\mu\Delta\tau)$ increases for larger delays $\Delta\tau$. But the increase of this difference is suppressed due to the triangular function appearing in (\ref{EQ:RF}) such that Equation~(\ref{EQ:tmax}) is a good approximation for the delay time at which the discrepancy (\ref{EQ:discApp}) is maximized. Inserting (\ref{EQ:tmax}) in (\ref{EQ:discApp}) and making a power series expansion around $z=0$ yields to first order in $z$
\begin{align}
\Delta P^c(\Delta\tau_{\max})=\frac{\mu}{4\xi}\sin\left(\frac{\mu}{2\xi}\right)z+\mathcal{O}(z^2).\label{EQ:PMax}
\end{align}
The occurrence of the sine-function in (\ref{EQ:PMax}) is rooted in the oscillating behavior of $\Delta P^c(\Delta\tau_{})$ as seen in Figure~\ref{fig:pattern}. We are only interested in the envelope of the difference $\Delta P^c(\Delta\tau_{})$ in Figure~\ref{fig:pattern} and therefore set the value of the sine-function in (\ref{EQ:PMax}) equal to one, i.e. we get
\begin{align}
\Delta P^c(\Delta\tau_{\max})\approx\frac{\mu}{4\xi}z.\label{EQ:PMax2}
\end{align}
We confirmed the validity of the approximation (\ref{EQ:PMax2}) numerically within an accuracy of less than $1\%$ for the parameter values for $\mu$ and $\xi$ shown in Figure~\ref{fig:Z}(a) and (b) under the condition $\mu/\xi > 1$ and $\Delta P^c(\Delta\tau_{\max})< 0.9$.

Equation (\ref{EQ:PMax2}) is a rough error estimation for technologies based on HOM-interference with frequency entangled photons, such as remote clock synchronization \citep{quan2016demonstration}, since it quantifies the influence of the redshift between the photon source and the measurement laboratory on the coincidence statistics.
 
From (\ref{EQ:PMax2}) we obtain an estimate for the ratio of the frequency separation to the single photon bandwidth $\mu/\xi$, which must be achieved in an experiment with a given detector resolution $\Delta P^c_{\mathrm{res}}$ in order to resolve the influence of a given redshift $z$. We find
\begin{align}
\frac{\mu}{\xi}=\frac{4 \Delta P^c_{\mathrm{res}}}{z}.\label{EQ:FEQ}
\end{align}  
Equation~(\ref{EQ:FEQ}) is a benchmark for future experiments, which intend to reveal the influence of the relativistic redshift on the coincidence probability of a HOM-experiment employing frequency entangled photons. For instance assuming a quite moderate detector resolution of $\Delta P^c_{\mathrm{res}}=0.1$ and a redshift between two counter rotating Earth satellites of $z=10^{-5}$ would amount for a frequency separation to single photon bandwidth ratio of $\mu/\xi=4\cdot10^{4}$. To our knowledge, the highest frequency separation $\mu\approx100\,\mathrm{THz}$ of frequency entangled photons has been already demonstrated \citep{chen2019hong}. There, a single photon bandwidth of $\xi\approx0.253\,\mathrm{THz}$ was achieved leading to a value $\mu/\xi\approx 4\cdot 10^2$, which could reveal the influence of the relativistic Doppler-shift between counter propagating satellites provided by a detector resolution of $1\%$, i.e. $\Delta P^c_{\mathrm{res}}=0.01$. No sub-GHz narrow band generation of frequency entangled photon pairs has been reported to date. In the case of cavity enhanced spontaneous parametric down conversion of frequency degenerated photons, ultra-narrowband emission with $\xi\approx265\,\mathrm{kHz}$ has also already been demonstrated \citep{liu2020sub}. In combination with a frequency separation $\mu\approx100\,\mathrm{THz}$, this would suffice to even resolve the influence of the general relativistic redshift of $z^{-10}$ between a geo-station on Earth and a satellite since these parameters satisfy Equation~(\ref{EQ:FEQ}) to a good approximation for a detector resolution of $\Delta P^c_{\mathrm{res}}=0.01$.

We want to place particular emphasis on the fact that the maximum difference between two HOM-interference experiments employing frequency entangled photons with and without relativistic aspects can reach unity in the limit $\mu/\xi\rightarrow\infty$ as can be seen in Figure~(\ref{fig:Z})(a) and (b). This means that, in the case of identical locally adjusted optical delays, the outcome of a coincidence measurement of frequency entangled photons can change between ideal photon bunching ($P^c=0$) to ideal photon anti-bunching ($P^c=1$) and vice versa as a function of relativistic redshift. It has been shown that the necessary and sufficient criterion for the occurrence of ideal photon bunching of two photons is that their joint spectral wave function is symmetric, i.e., $\phi(\omega_1,\omega_2)=\phi(\omega_2,\omega_1)$, see \citep{HOMTheory}. In that case, it was also shown that the necessary and sufficient criterion for the occurrence of ideal photon anti-bunching of two photons is that their joint spectral wave function is anti-symmetric, i.e., $\phi(\omega_1,\omega_2)=-\phi(\omega_2,\omega_1)$. In the case of quantum beating, i.e. a periodic change between photon bunching and anti-bunching in the delay, the authors made the interesting argument that due to the adjustment of different optical delays the parity of the spectral wave function changes from being symmetric to anti-symmetric and vice versa. This means that one observer might observe photon bunching thereby exploring a symmetric spectral wave function with a certain adjustment of the local optical delays, while another observer with mutual redshift with respect to the first one might explore an anti-symmetric wave function through the measurement of photon anti-bunching with the same setting of the (local) optical delays in his local laboratory. From this we infer that the parity of the joint spectral wave function is observer dependent. The peculiarity of our results is that, the parity of the joint spectral wave function does not only depend on the observer's delay settings but also on his state of motion. This is a central result of our paper. This observer dependence becomes especially important in the case of spectrally indistinguishable frequency entangled photons in the limit of large frequency separations. However, according to Equation~(\ref{EQ:relAB}) the coincidence detection probability at the zero delay setting (i.e. $\Delta\tau=\tau_1-\tau_2=0$) is independent of the redshift between the photon source and the measurement laboratory and thus can yield some information about the optical paths through which the photons propagated before they interfere at the common BS of the detection instance (see Figure~\ref{fig:RelHOM}).

Finally we discuss the behavior of frequency-detuned quantum states (see Equation~(\ref{EQ:FE1}) for spectral profile).
This case is less interesting than the one of frequency entangled photons. Here the discrepancy between the coincidence detection probabilities reads
\begin{align}
\Delta P^c(\Delta\tau)=\frac{1}{2}(S_{\mu\xi}((1+z)\Delta\tau)-S_{\mu\xi}(\Delta\tau)),
\end{align}
with $S_{\mu\xi}(\tau)$ from (\ref{EQ:SF}).
Figure~\ref{fig:Z}(c) and (d) show the absolute value of the maximum discrepancy of a HOM-experiment, which uses frequency detuned photons under the influence of a redshift of $z=10^{-10}$ and $z=10^{-5}$ with respect to a HOM-experiment without relativistic influence.
It can be seen that in the limit $\mu/\xi\rightarrow 0$ the maximum discrepancy is $|\Delta P^c(\Delta\tau_{\max})|\approx z/2$, which cannot be resolved in a HOM-experiment for realistic redshifts $z\lessapprox10^{-5}$. Taking the limit $\mu/\xi\rightarrow \infty$ even suppresses the effect further. This is originated in the fact that in this limit $S_{\mu\xi}(\tau)$ and thus $\Delta P^c(\Delta\tau)$ approaches zero, which by itself is rooted in the increasing spectral distinguishability of the frequency detuned photons. Since spectrally indistinguishable frequency entangled photons do not suffer from this effect, makes them more attractive to study the impact of general relativity on quantum mechanics.

\section{\label{sec:outlook} Outlook}
There are several possible extensions to this work.
The presented framework can be used to study the spectral aspects of Sagnac-interferometry \citep{Sagnac}, where photons impact from different directions on moving BSs to interfere. While the present work is concerned with theoretical and fundamental aspects of relativistic and quantum photonic science, it is also of great interest to study relativistic effects in HOM-interference-based quantum technologies. For example, one can quantify the change in accuracy within HOM-based space-based clock synchronization \citep{giovannetti2001quantum}. Furthermore, a greater understanding of the effects described here and in previous work \cite{Bruschi1,Bruschi2,Bruschi3,Bruschi4} would be achieved by considering different initial states, such as coherent states, squeezed states, cat states and, more realistically, mixed states. Mixed states in particular are expected to provide a benchmark for the effects witnessed by pure states as it was already shown in the context of gravitational redshift effects on quantum states \citep{Bruschi3}. Particular attention can also be given to optimizing spectral wave functions and the initial quantum states to obtain to greatest (or the least) gravitational effects on HOM-interference.

Another aspect to be investigated is gravity-induced entanglement dynamics of photons that propagate through different (possibly correlated) paths in spacetime. Intriguingly recent works showed that gravitational effects give rise to change the entanglement between different light field modes, which can be envisaged in two-photon interference experiments such as the HOM-experiment \citep{Bruschi4}. Other work claims that gravitational time dilation induces entanglement between external and internal DOFs of quantum particles \citep{zych2011quantum}. We think that HOM-experiments can be interpreted precisely in this way. The occurrence of quantum-beating would correspond to a periodic creation and depletion of entanglement between the internal and external DOFs, complementary to the depletion and creation of frequency entanglement among the photons. We believe that our formalism can be extended to study both of these aspects and yield a more comprehensive understanding on the influence of general relativity on genuine quantum features of physical systems.

\section{\label{sec:conclusions} Conclusions}
We have employed Glauber's theory of optical coherence, together with a quantum field theoretical framework previously developed to describe the deformation of the frequency profile of photons propagating through curved spacetime \citep{Bruschi1,Bruschi2}, to develop a formalism that describes large-scale HOM-experiments where quantum and relativistic effects are both non-negligible.

In HOM-experiments the particle nature of photons plays a special role. Such experiments can only be explained by means of quantum mechanical principles such as the indistinguishability and quantum entanglement of the involved systems. These lead to phenomena like photon bunching and photon anti-bunching, both without classical analogue. Therefore, our predictions of the effects of the gravitational frequency shift on these inherently quantum mechanical experiments can be interpreted as a genuine quantum test of gravity. Apart from the fundamental aspects of our work, the results presented here are also of practical relevance since they provide methods to estimate how strongly quantum technological applications based on HOM-interference might be distorted from an ideal operation (see Equation~(\ref{EQ:FEQ})).

We focussed our methods on three cases with different photon sources, each one generating photons with a different spectral profile: the case analogous to the one found in the original HOM-experiment \citep{HOM}, which employs spectrally indistinguishable frequency uncorrelated photons; the case of spectrally distinguishable frequency detuned photons \citep{Zeillinger2}; the case of spectrally indistinguishable frequency entangled photons \citep{Zeillinger}. We found that in the original HOM-experiment no adjustment of the experimental parameters (which consisted in the photon's bandwidth alone) yields an amplification of the relativistic impact on HOM-interferometry. This has the interesting implication that in case of frequency uncorrelated photons the impact of the redshift in HOM-interference is universal. Concretely, this means that it depends only on the redshift between the photon source and the detectors, and not on the photon's bandwidth (as can be seen from Equation~(\ref{EQ:Pmax1})).
This contrasts with the case of frequency entangled photons. In this case we have shown that, due to the increase of the photon frequency separation $\mu$ and decrease of the single photon bandwidth $\xi$, the effect of the relativistic frequency shift becomes more important since higher frequency separations lead to an increasing sensitivity of the occurrence of photon bunching or anti-bunching on the delay. A particularly interesting property of frequency entangled photons is that the relativistic frequency shift between the photon source and the detectors can change the detection statistics from ideal photon bunching to ideal photon anti-bunching, and vice versa. This is reflected in the fact that Equation~(\ref{EQ:discApp}) can reach unity. Since photon bunching in quantum mechanics is equivalent to the symmetry of the joint spectral wave function, and photon anti-bunching is equivalent to the anti-symmetry of the joint spectral wave function \citep{HOMTheory}, this leads to conclude that the parity of the joint photonic spectral wave function is an observer dependent quantity, which is one of the central results of this work. We conclude from this that the outcomes of HOM-experiments rely not only on the optical path difference between the measured photons before detection, but also on the state of motion of the photon source and the observers.

The main contribution of the present work consists in a formalism that can be used to describe the impact of the relativistic redshift on the spectral aspects of quantum entanglement and quantum indistinguishability of photons within the context of HOM-interference. Our work, therefore, adds to the ongoing effort of understanding the interplay between gravity and quantum coherence and entanglement \cite{feynman2018feynman,Bruschi3}, and constitutes another step towards the development of  relativistic and quantum technologies. Experimental verification of the effects presented here, as well as in this body of work more broadly, can serve as a demonstration of quantum field theory in (weakly) curved spacetime, and provide new insights in the quest for a unified theory of Nature.

\acknowledgments
R.B. and A.W.S. were funded by the Deutsche Forschungsgemeinschaft (DFG, German Research Foundation) under Germany’s Excellence Strategy – EXC-2123 QuantumFrontiers – 390837967. We further gratefully acknowledge support through the TerraQ initiative from the Deutsche Forschungsgemeinschaft (DFG, German Research Foundation) – Project-ID 434617780 – SFB 1464 and the Research Training Group 1620 “Models of Gravity”.
\bibliographystyle{apsrev4-2}
\bibliography{bibtestA}

\begin{thebibliography}{42}%
\makeatletter
\providecommand \@ifxundefined [1]{%
 \@ifx{#1\undefined}
}%
\providecommand \@ifnum [1]{%
 \ifnum #1\expandafter \@firstoftwo
 \else \expandafter \@secondoftwo
 \fi
}%
\providecommand \@ifx [1]{%
 \ifx #1\expandafter \@firstoftwo
 \else \expandafter \@secondoftwo
 \fi
}%
\providecommand \natexlab [1]{#1}%
\providecommand \enquote  [1]{``#1''}%
\providecommand \bibnamefont  [1]{#1}%
\providecommand \bibfnamefont [1]{#1}%
\providecommand \citenamefont [1]{#1}%
\providecommand \href@noop [0]{\@secondoftwo}%
\providecommand \href [0]{\begingroup \@sanitize@url \@href}%
\providecommand \@href[1]{\@@startlink{#1}\@@href}%
\providecommand \@@href[1]{\endgroup#1\@@endlink}%
\providecommand \@sanitize@url [0]{\catcode `\\12\catcode `\$12\catcode
  `\&12\catcode `\#12\catcode `\^12\catcode `\_12\catcode `\%12\relax}%
\providecommand \@@startlink[1]{}%
\providecommand \@@endlink[0]{}%
\providecommand \url  [0]{\begingroup\@sanitize@url \@url }%
\providecommand \@url [1]{\endgroup\@href {#1}{\urlprefix }}%
\providecommand \urlprefix  [0]{URL }%
\providecommand \Eprint [0]{\href }%
\providecommand \doibase [0]{https://doi.org/}%
\providecommand \selectlanguage [0]{\@gobble}%
\providecommand \bibinfo  [0]{\@secondoftwo}%
\providecommand \bibfield  [0]{\@secondoftwo}%
\providecommand \translation [1]{[#1]}%
\providecommand \BibitemOpen [0]{}%
\providecommand \bibitemStop [0]{}%
\providecommand \bibitemNoStop [0]{.\EOS\space}%
\providecommand \EOS [0]{\spacefactor3000\relax}%
\providecommand \BibitemShut  [1]{\csname bibitem#1\endcsname}%
\let\auto@bib@innerbib\@empty
\bibitem [{\citenamefont {Feynman}\ \emph {et~al.}(2018)\citenamefont
  {Feynman}, \citenamefont {Morinigo}, \citenamefont {Wagner}, \citenamefont
  {Hatfield}, \citenamefont {Preskill},\ and\ \citenamefont
  {Thorne}}]{feynman2018feynman}%
  \BibitemOpen
  \bibfield  {author} {\bibinfo {author} {\bibfnamefont {R.}~\bibnamefont
  {Feynman}}, \bibinfo {author} {\bibfnamefont {F.}~\bibnamefont {Morinigo}},
  \bibinfo {author} {\bibfnamefont {W.}~\bibnamefont {Wagner}}, \bibinfo
  {author} {\bibfnamefont {B.}~\bibnamefont {Hatfield}}, \bibinfo {author}
  {\bibfnamefont {J.}~\bibnamefont {Preskill}},\ and\ \bibinfo {author}
  {\bibfnamefont {K.}~\bibnamefont {Thorne}},\ }\href@noop {} {\emph {\bibinfo
  {title} {Feynman lectures on gravitation}}}\ (\bibinfo  {publisher} {CRC
  Press},\ \bibinfo {year} {2018})\BibitemShut {NoStop}%
\bibitem [{\citenamefont {Colella}\ \emph {et~al.}(1975)\citenamefont
  {Colella}, \citenamefont {Overhauser},\ and\ \citenamefont {Werner}}]{COW}%
  \BibitemOpen
  \bibfield  {author} {\bibinfo {author} {\bibfnamefont {R.}~\bibnamefont
  {Colella}}, \bibinfo {author} {\bibfnamefont {A.~W.}\ \bibnamefont
  {Overhauser}},\ and\ \bibinfo {author} {\bibfnamefont {S.~A.}\ \bibnamefont
  {Werner}},\ }\href {https://doi.org/10.1103/PhysRevLett.34.1472} {\bibfield
  {journal} {\bibinfo  {journal} {Phys. Rev. Lett.}\ }\textbf {\bibinfo
  {volume} {34}},\ \bibinfo {pages} {1472} (\bibinfo {year}
  {1975})}\BibitemShut {NoStop}%
\bibitem [{\citenamefont {Pound}\ and\ \citenamefont
  {Rebka}(1960)}]{PhotRedshift}%
  \BibitemOpen
  \bibfield  {author} {\bibinfo {author} {\bibfnamefont {R.}~\bibnamefont
  {Pound}}\ and\ \bibinfo {author} {\bibfnamefont {G.}~\bibnamefont {Rebka}},\
  }\href {https://doi.org/10.1103/PhysRevLett.4.337} {\bibfield  {journal}
  {\bibinfo  {journal} {Phys. Rev. Lett.}\ }\textbf {\bibinfo {volume} {4}},\
  \bibinfo {pages} {337} (\bibinfo {year} {1960})}\BibitemShut {NoStop}%
\bibitem [{\citenamefont {Pound}\ and\ \citenamefont
  {Snider}(1964)}]{PhotRedshift2}%
  \BibitemOpen
  \bibfield  {author} {\bibinfo {author} {\bibfnamefont {R.}~\bibnamefont
  {Pound}}\ and\ \bibinfo {author} {\bibfnamefont {J.}~\bibnamefont {Snider}},\
  }\href {https://doi.org/10.1103/PhysRevLett.13.539} {\bibfield  {journal}
  {\bibinfo  {journal} {Phys. Rev. Lett.}\ }\textbf {\bibinfo {volume} {13}},\
  \bibinfo {pages} {539} (\bibinfo {year} {1964})}\BibitemShut {NoStop}%
\bibitem [{\citenamefont {M{\"u}ller}\ \emph {et~al.}(2010)\citenamefont
  {M{\"u}ller}, \citenamefont {Peters},\ and\ \citenamefont
  {Chu}}]{Mueller:Peters:2010}%
  \BibitemOpen
  \bibfield  {author} {\bibinfo {author} {\bibfnamefont {H.}~\bibnamefont
  {M{\"u}ller}}, \bibinfo {author} {\bibfnamefont {A.}~\bibnamefont {Peters}},\
  and\ \bibinfo {author} {\bibfnamefont {S.}~\bibnamefont {Chu}},\ }\href
  {https://doi.org/10.1038/nature08776} {\bibfield  {journal} {\bibinfo
  {journal} {Nature}\ }\textbf {\bibinfo {volume} {463}},\ \bibinfo {pages}
  {926} (\bibinfo {year} {2010})}\BibitemShut {NoStop}%
\bibitem [{\citenamefont {Dowling}\ and\ \citenamefont
  {Milburn}(2003)}]{SecondQuantumRevolution}%
  \BibitemOpen
  \bibfield  {author} {\bibinfo {author} {\bibfnamefont {J.}~\bibnamefont
  {Dowling}}\ and\ \bibinfo {author} {\bibfnamefont {G.~J.}\ \bibnamefont
  {Milburn}},\ }\href@noop {} {\bibfield  {journal} {\bibinfo  {journal}
  {Philosophical Transactions of the Royal Society of London. Series A:
  Mathematical, Physical and Engineering Sciences}\ }\textbf {\bibinfo {volume}
  {361}},\ \bibinfo {pages} {1655} (\bibinfo {year} {2003})}\BibitemShut
  {NoStop}%
\bibitem [{\citenamefont {Yin}\ \emph {et~al.}(2017{\natexlab{a}})\citenamefont
  {Yin}, \citenamefont {Cao}, \citenamefont {Li}, \citenamefont {Liao},
  \citenamefont {Zhang}, \citenamefont {Ren}, \citenamefont {Cai},
  \citenamefont {Liu}, \citenamefont {Li},\ and\ \citenamefont
  {\textit{et.al.}}}]{1200}%
  \BibitemOpen
  \bibfield  {author} {\bibinfo {author} {\bibfnamefont {J.}~\bibnamefont
  {Yin}}, \bibinfo {author} {\bibfnamefont {Y.}~\bibnamefont {Cao}}, \bibinfo
  {author} {\bibfnamefont {Y.-H.}\ \bibnamefont {Li}}, \bibinfo {author}
  {\bibfnamefont {S.-K.}\ \bibnamefont {Liao}}, \bibinfo {author}
  {\bibfnamefont {L.}~\bibnamefont {Zhang}}, \bibinfo {author} {\bibfnamefont
  {J.-G.}\ \bibnamefont {Ren}}, \bibinfo {author} {\bibfnamefont {W.-Q.}\
  \bibnamefont {Cai}}, \bibinfo {author} {\bibfnamefont {W.-Y.}\ \bibnamefont
  {Liu}}, \bibinfo {author} {\bibfnamefont {B.}~\bibnamefont {Li}},\ and\
  \bibinfo {author} {\bibfnamefont {H.~D.}\ \bibnamefont {\textit{et.al.}}},\
  }\href {https://doi.org/10.1126/science.aan3211} {\bibfield  {journal}
  {\bibinfo  {journal} {Science}\ }\textbf {\bibinfo {volume} {356}},\ \bibinfo
  {pages} {1140} (\bibinfo {year} {2017}{\natexlab{a}})},\ \Eprint
  {https://arxiv.org/abs/https://www.science.org/doi/pdf/10.1126/science.aan3211}
  {https://www.science.org/doi/pdf/10.1126/science.aan3211} \BibitemShut
  {NoStop}%
\bibitem [{\citenamefont {Liao}\ \emph
  {et~al.}(2017{\natexlab{a}})\citenamefont {Liao}, \citenamefont {Cai},
  \citenamefont {Liu}, \citenamefont {Zhang}, \citenamefont {Li}, \citenamefont
  {Ren}, \citenamefont {Yin}, \citenamefont {Shen}, \citenamefont {Cao},\ and\
  \citenamefont {\textit{et.al.}}}]{liao2017satellite}%
  \BibitemOpen
  \bibfield  {author} {\bibinfo {author} {\bibfnamefont {S.}~\bibnamefont
  {Liao}}, \bibinfo {author} {\bibfnamefont {W.}~\bibnamefont {Cai}}, \bibinfo
  {author} {\bibfnamefont {W.}~\bibnamefont {Liu}}, \bibinfo {author}
  {\bibfnamefont {L.}~\bibnamefont {Zhang}}, \bibinfo {author} {\bibfnamefont
  {Y.}~\bibnamefont {Li}}, \bibinfo {author} {\bibfnamefont {J.}~\bibnamefont
  {Ren}}, \bibinfo {author} {\bibfnamefont {J.}~\bibnamefont {Yin}}, \bibinfo
  {author} {\bibfnamefont {Q.}~\bibnamefont {Shen}}, \bibinfo {author}
  {\bibfnamefont {Y.}~\bibnamefont {Cao}},\ and\ \bibinfo {author}
  {\bibfnamefont {Z.~L.}\ \bibnamefont {\textit{et.al.}}},\ }\href@noop {}
  {\bibfield  {journal} {\bibinfo  {journal} {Nature}\ }\textbf {\bibinfo
  {volume} {549}},\ \bibinfo {pages} {43} (\bibinfo {year}
  {2017}{\natexlab{a}})}\BibitemShut {NoStop}%
\bibitem [{\citenamefont {Ren}\ \emph {et~al.}(2017)\citenamefont {Ren},
  \citenamefont {Xu}, \citenamefont {Yong}, \citenamefont {Zhang},
  \citenamefont {Liao}, \citenamefont {Yin}, \citenamefont {Liu}, \citenamefont
  {Cai}, \citenamefont {Yang},\ and\ \citenamefont {L.~Li}}]{ren2017ground}%
  \BibitemOpen
  \bibfield  {author} {\bibinfo {author} {\bibfnamefont {J.}~\bibnamefont
  {Ren}}, \bibinfo {author} {\bibfnamefont {P.}~\bibnamefont {Xu}}, \bibinfo
  {author} {\bibfnamefont {H.}~\bibnamefont {Yong}}, \bibinfo {author}
  {\bibfnamefont {L.}~\bibnamefont {Zhang}}, \bibinfo {author} {\bibfnamefont
  {S.}~\bibnamefont {Liao}}, \bibinfo {author} {\bibfnamefont {J.}~\bibnamefont
  {Yin}}, \bibinfo {author} {\bibfnamefont {W.}~\bibnamefont {Liu}}, \bibinfo
  {author} {\bibfnamefont {W.}~\bibnamefont {Cai}}, \bibinfo {author}
  {\bibfnamefont {M.}~\bibnamefont {Yang}},\ and\ \bibinfo {author}
  {\bibfnamefont {t.}~\bibnamefont {L.~Li}},\ }\href@noop {} {\bibfield
  {journal} {\bibinfo  {journal} {Nature}\ }\textbf {\bibinfo {volume} {549}},\
  \bibinfo {pages} {70} (\bibinfo {year} {2017})}\BibitemShut {NoStop}%
\bibitem [{\citenamefont {Yin}\ \emph {et~al.}(2017{\natexlab{b}})\citenamefont
  {Yin}, \citenamefont {Cao}, \citenamefont {Li}, \citenamefont {Ren},
  \citenamefont {Liao}, \citenamefont {Zhang}, \citenamefont {Cai},
  \citenamefont {Liu},\ and\ \citenamefont
  {\textit{et.al.}}}]{yin2017satellite}%
  \BibitemOpen
  \bibfield  {author} {\bibinfo {author} {\bibfnamefont {J.}~\bibnamefont
  {Yin}}, \bibinfo {author} {\bibfnamefont {Y.}~\bibnamefont {Cao}}, \bibinfo
  {author} {\bibfnamefont {Y.}~\bibnamefont {Li}}, \bibinfo {author}
  {\bibfnamefont {J.}~\bibnamefont {Ren}}, \bibinfo {author} {\bibfnamefont
  {S.}~\bibnamefont {Liao}}, \bibinfo {author} {\bibfnamefont {L.}~\bibnamefont
  {Zhang}}, \bibinfo {author} {\bibfnamefont {W.}~\bibnamefont {Cai}}, \bibinfo
  {author} {\bibfnamefont {W.}~\bibnamefont {Liu}},\ and\ \bibinfo {author}
  {\bibfnamefont {B.~L.}\ \bibnamefont {\textit{et.al.}}},\ }\href
  {https://doi.org/10.1103/PhysRevLett.119.200501} {\bibfield  {journal}
  {\bibinfo  {journal} {Phys. Rev. Lett.}\ }\textbf {\bibinfo {volume} {119}},\
  \bibinfo {pages} {200501} (\bibinfo {year} {2017}{\natexlab{b}})}\BibitemShut
  {NoStop}%
\bibitem [{\citenamefont {Liao}\ \emph
  {et~al.}(2017{\natexlab{b}})\citenamefont {Liao}, \citenamefont {Lin},
  \citenamefont {Ren}, \citenamefont {Liu}, \citenamefont {J.Qiang},
  \citenamefont {Yin}, \citenamefont {Li}, \citenamefont {Shen}, ,\ and\
  \citenamefont {Zhang}}]{liao2017space}%
  \BibitemOpen
  \bibfield  {author} {\bibinfo {author} {\bibfnamefont {S.}~\bibnamefont
  {Liao}}, \bibinfo {author} {\bibfnamefont {J.}~\bibnamefont {Lin}}, \bibinfo
  {author} {\bibfnamefont {J.}~\bibnamefont {Ren}}, \bibinfo {author}
  {\bibfnamefont {W.}~\bibnamefont {Liu}}, \bibinfo {author} {\bibnamefont
  {J.Qiang}}, \bibinfo {author} {\bibfnamefont {J.}~\bibnamefont {Yin}},
  \bibinfo {author} {\bibfnamefont {Y.}~\bibnamefont {Li}}, \bibinfo {author}
  {\bibfnamefont {Q.}~\bibnamefont {Shen}}, ,\ and\ \bibinfo {author}
  {\bibfnamefont {L.}~\bibnamefont {Zhang}},\ }\href
  {https://doi.org/10.1088/0256-307x/34/9/090302} {\bibfield  {journal}
  {\bibinfo  {journal} {Chinese Physics Letters}\ }\textbf {\bibinfo {volume}
  {34}},\ \bibinfo {pages} {090302} (\bibinfo {year}
  {2017}{\natexlab{b}})}\BibitemShut {NoStop}%
\bibitem [{\citenamefont {Liao}\ \emph {et~al.}(2018)\citenamefont {Liao},
  \citenamefont {Cai}, \citenamefont {Handsteiner}, \citenamefont {Liu},
  \citenamefont {Yin}, \citenamefont {Zhang}, \citenamefont {Rauch},
  \citenamefont {Fink}, \citenamefont {Ren},\ and\ \citenamefont
  {\textit{et.al.}}}]{liao2018satellite}%
  \BibitemOpen
  \bibfield  {author} {\bibinfo {author} {\bibfnamefont {S.}~\bibnamefont
  {Liao}}, \bibinfo {author} {\bibfnamefont {W.}~\bibnamefont {Cai}}, \bibinfo
  {author} {\bibfnamefont {J.}~\bibnamefont {Handsteiner}}, \bibinfo {author}
  {\bibfnamefont {B.}~\bibnamefont {Liu}}, \bibinfo {author} {\bibfnamefont
  {J.}~\bibnamefont {Yin}}, \bibinfo {author} {\bibfnamefont {L.}~\bibnamefont
  {Zhang}}, \bibinfo {author} {\bibfnamefont {D.}~\bibnamefont {Rauch}},
  \bibinfo {author} {\bibfnamefont {M.}~\bibnamefont {Fink}}, \bibinfo {author}
  {\bibfnamefont {G.}~\bibnamefont {Ren}},\ and\ \bibinfo {author}
  {\bibfnamefont {W.~L.}\ \bibnamefont {\textit{et.al.}}},\ }\href
  {https://doi.org/10.1103/PhysRevLett.120.030501} {\bibfield  {journal}
  {\bibinfo  {journal} {Phys. Rev. Lett.}\ }\textbf {\bibinfo {volume} {120}},\
  \bibinfo {pages} {030501} (\bibinfo {year} {2018})}\BibitemShut {NoStop}%
\bibitem [{\citenamefont {Mann}\ and\ \citenamefont
  {Ralph}(2012)}]{Mann:Ralph:2012}%
  \BibitemOpen
  \bibfield  {author} {\bibinfo {author} {\bibfnamefont {R.~B.}\ \bibnamefont
  {Mann}}\ and\ \bibinfo {author} {\bibfnamefont {T.~C.}\ \bibnamefont
  {Ralph}},\ }\href {https://doi.org/10.1088/0264-9381/29/22/220301} {\bibfield
   {journal} {\bibinfo  {journal} {Classical and Quantum Gravity}\ }\textbf
  {\bibinfo {volume} {29}},\ \bibinfo {pages} {220301} (\bibinfo {year}
  {2012})}\BibitemShut {NoStop}%
\bibitem [{\citenamefont {Bruschi}\ \emph
  {et~al.}(2014{\natexlab{a}})\citenamefont {Bruschi}, \citenamefont {Ralph},
  \citenamefont {Fuentes}, \citenamefont {Jennewein},\ and\ \citenamefont
  {Razavi}}]{Bruschi1}%
  \BibitemOpen
  \bibfield  {author} {\bibinfo {author} {\bibfnamefont {D.}~\bibnamefont
  {Bruschi}}, \bibinfo {author} {\bibfnamefont {T.}~\bibnamefont {Ralph}},
  \bibinfo {author} {\bibfnamefont {I.}~\bibnamefont {Fuentes}}, \bibinfo
  {author} {\bibfnamefont {T.}~\bibnamefont {Jennewein}},\ and\ \bibinfo
  {author} {\bibfnamefont {M.}~\bibnamefont {Razavi}},\ }\href
  {https://doi.org/10.1103/PhysRevD.90.045041} {\bibfield  {journal} {\bibinfo
  {journal} {Phys. Rev. D}\ }\textbf {\bibinfo {volume} {90}},\ \bibinfo
  {pages} {045041} (\bibinfo {year} {2014}{\natexlab{a}})}\BibitemShut
  {NoStop}%
\bibitem [{\citenamefont {Bruschi}\ \emph
  {et~al.}(2014{\natexlab{b}})\citenamefont {Bruschi}, \citenamefont {Datta},
  \citenamefont {Ursin}, \citenamefont {Ralph},\ and\ \citenamefont
  {Fuentes}}]{Bruschi2}%
  \BibitemOpen
  \bibfield  {author} {\bibinfo {author} {\bibfnamefont {D.}~\bibnamefont
  {Bruschi}}, \bibinfo {author} {\bibfnamefont {A.}~\bibnamefont {Datta}},
  \bibinfo {author} {\bibfnamefont {R.}~\bibnamefont {Ursin}}, \bibinfo
  {author} {\bibfnamefont {T.}~\bibnamefont {Ralph}},\ and\ \bibinfo {author}
  {\bibfnamefont {I.}~\bibnamefont {Fuentes}},\ }\href
  {https://doi.org/10.1103/PhysRevD.90.124001} {\bibfield  {journal} {\bibinfo
  {journal} {Phys. Rev. D}\ }\textbf {\bibinfo {volume} {90}},\ \bibinfo
  {pages} {124001} (\bibinfo {year} {2014}{\natexlab{b}})}\BibitemShut
  {NoStop}%
\bibitem [{\citenamefont {Bruschi}\ and\ \citenamefont
  {Schell}(2021)}]{Bruschi4}%
  \BibitemOpen
  \bibfield  {author} {\bibinfo {author} {\bibfnamefont {D.~E.}\ \bibnamefont
  {Bruschi}}\ and\ \bibinfo {author} {\bibfnamefont {A.~W.}\ \bibnamefont
  {Schell}},\ }\href@noop {} {\  (\bibinfo {year} {2021})},\ \Eprint
  {https://arxiv.org/abs/2109.00728} {arXiv:2109.00728 [quant-ph]} \BibitemShut
  {NoStop}%
\bibitem [{\citenamefont {Schnabel}(2017)}]{schnabel2017squeezed}%
  \BibitemOpen
  \bibfield  {author} {\bibinfo {author} {\bibfnamefont {R.}~\bibnamefont
  {Schnabel}},\ }\href
  {https://doi.org/https://doi.org/10.1016/j.physrep.2017.04.001} {\bibfield
  {journal} {\bibinfo  {journal} {Phys. Rep.}\ }\textbf {\bibinfo {volume}
  {684}},\ \bibinfo {pages} {1} (\bibinfo {year} {2017})},\ \bibinfo {note}
  {squeezed states of light and their applications in laser
  interferometers}\BibitemShut {NoStop}%
\bibitem [{\citenamefont {Giovannetti}\ \emph {et~al.}(2001)\citenamefont
  {Giovannetti}, \citenamefont {Lloyd},\ and\ \citenamefont
  {Maccone}}]{giovannetti2001quantum}%
  \BibitemOpen
  \bibfield  {author} {\bibinfo {author} {\bibfnamefont {V.}~\bibnamefont
  {Giovannetti}}, \bibinfo {author} {\bibfnamefont {S.}~\bibnamefont {Lloyd}},\
  and\ \bibinfo {author} {\bibfnamefont {L.}~\bibnamefont {Maccone}},\
  }\href@noop {} {\bibfield  {journal} {\bibinfo  {journal} {Nature}\ }\textbf
  {\bibinfo {volume} {412}},\ \bibinfo {pages} {417} (\bibinfo {year}
  {2001})}\BibitemShut {NoStop}%
\bibitem [{\citenamefont {Giovannetti}\ and\ \citenamefont
  {Lloyd}(2004)}]{giovannetti2004quantum}%
  \BibitemOpen
  \bibfield  {author} {\bibinfo {author} {\bibfnamefont {V.}~\bibnamefont
  {Giovannetti}}\ and\ \bibinfo {author} {\bibfnamefont {L.}~\bibnamefont
  {Lloyd}, \bibfnamefont {S.and~Maccone}},\ }\href@noop {} {\bibfield
  {journal} {\bibinfo  {journal} {Science}\ }\textbf {\bibinfo {volume}
  {306}},\ \bibinfo {pages} {1330} (\bibinfo {year} {2004})}\BibitemShut
  {NoStop}%
\bibitem [{\citenamefont {Hong}\ \emph {et~al.}(1987)\citenamefont {Hong},
  \citenamefont {Ou},\ and\ \citenamefont {Mandel}}]{HOM}%
  \BibitemOpen
  \bibfield  {author} {\bibinfo {author} {\bibfnamefont {C.}~\bibnamefont
  {Hong}}, \bibinfo {author} {\bibfnamefont {Z.}~\bibnamefont {Ou}},\ and\
  \bibinfo {author} {\bibfnamefont {L.}~\bibnamefont {Mandel}},\ }\href
  {https://doi.org/10.1103/PhysRevLett.59.2044} {\bibfield  {journal} {\bibinfo
   {journal} {Phys. Rev. Lett.}\ }\textbf {\bibinfo {volume} {59}},\ \bibinfo
  {pages} {2044} (\bibinfo {year} {1987})}\BibitemShut {NoStop}%
\bibitem [{\citenamefont {Quan}\ \emph {et~al.}(2016)\citenamefont {Quan},
  \citenamefont {Zhai}, \citenamefont {Wang}, \citenamefont {Hou},
  \citenamefont {Wang}, \citenamefont {Xiang}, \citenamefont {Liu},
  \citenamefont {Zhang},\ and\ \citenamefont {Dong}}]{quan2016demonstration}%
  \BibitemOpen
  \bibfield  {author} {\bibinfo {author} {\bibfnamefont {R.}~\bibnamefont
  {Quan}}, \bibinfo {author} {\bibfnamefont {Y.}~\bibnamefont {Zhai}}, \bibinfo
  {author} {\bibfnamefont {M.}~\bibnamefont {Wang}}, \bibinfo {author}
  {\bibfnamefont {F.}~\bibnamefont {Hou}}, \bibinfo {author} {\bibfnamefont
  {S.}~\bibnamefont {Wang}}, \bibinfo {author} {\bibfnamefont {X.}~\bibnamefont
  {Xiang}}, \bibinfo {author} {\bibfnamefont {T.}~\bibnamefont {Liu}}, \bibinfo
  {author} {\bibfnamefont {S.}~\bibnamefont {Zhang}},\ and\ \bibinfo {author}
  {\bibfnamefont {R.}~\bibnamefont {Dong}},\ }\href@noop {} {\bibfield
  {journal} {\bibinfo  {journal} {Scientific reports}\ }\textbf {\bibinfo
  {volume} {6}},\ \bibinfo {pages} {1} (\bibinfo {year} {2016})}\BibitemShut
  {NoStop}%
\bibitem [{\citenamefont {Valencia}\ and\ \citenamefont
  {Scarcelli}(2004)}]{valencia2004distant}%
  \BibitemOpen
  \bibfield  {author} {\bibinfo {author} {\bibfnamefont {A.}~\bibnamefont
  {Valencia}}\ and\ \bibinfo {author} {\bibfnamefont {Y.}~\bibnamefont
  {Scarcelli}, \bibfnamefont {G.and~Shih}},\ }\href@noop {} {\bibfield
  {journal} {\bibinfo  {journal} {Appl. Phys. Lett.}\ }\textbf {\bibinfo
  {volume} {85}},\ \bibinfo {pages} {2655} (\bibinfo {year}
  {2004})}\BibitemShut {NoStop}%
\bibitem [{\citenamefont {Barzel}\ and\ \citenamefont
  {Laemmerzahl}(2022)}]{Barzel2}%
  \BibitemOpen
  \bibfield  {author} {\bibinfo {author} {\bibfnamefont {R.}~\bibnamefont
  {Barzel}}\ and\ \bibinfo {author} {\bibfnamefont {C.}~\bibnamefont
  {Laemmerzahl}},\ }\href@noop {} {\  (\bibinfo {year} {2022})},\ \Eprint
  {https://arxiv.org/abs/2202.07522} {arXiv:2202.07522 [quant-ph]} \BibitemShut
  {NoStop}%
\bibitem [{\citenamefont {Pradana}\ and\ \citenamefont
  {Chew}(2019)}]{pradana2019quantum}%
  \BibitemOpen
  \bibfield  {author} {\bibinfo {author} {\bibfnamefont {A.}~\bibnamefont
  {Pradana}}\ and\ \bibinfo {author} {\bibfnamefont {L.}~\bibnamefont {Chew}},\
  }\href {https://doi.org/10.1088/1367-2630/ab1bbf} {\bibfield  {journal}
  {\bibinfo  {journal} {New J. Phys.}\ }\textbf {\bibinfo {volume} {21}},\
  \bibinfo {pages} {053027} (\bibinfo {year} {2019})}\BibitemShut {NoStop}%
\bibitem [{\citenamefont {Wang}(2006)}]{HOMTheory}%
  \BibitemOpen
  \bibfield  {author} {\bibinfo {author} {\bibfnamefont {K.}~\bibnamefont
  {Wang}},\ }\href {https://doi.org/10.1088/0953-4075/39/18/r01} {\bibfield
  {journal} {\bibinfo  {journal} {J. Phys. B: At. Mol. Opt. Phys.}\ }\textbf
  {\bibinfo {volume} {39}},\ \bibinfo {pages} {R293} (\bibinfo {year}
  {2006})}\BibitemShut {NoStop}%
\bibitem [{\citenamefont {Ou}\ and\ \citenamefont
  {Mandel}(1988)}]{ou1988observation}%
  \BibitemOpen
  \bibfield  {author} {\bibinfo {author} {\bibfnamefont {Z.}~\bibnamefont
  {Ou}}\ and\ \bibinfo {author} {\bibfnamefont {L.}~\bibnamefont {Mandel}},\
  }\href {https://doi.org/10.1103/PhysRevLett.61.54} {\bibfield  {journal}
  {\bibinfo  {journal} {Phys. Rev. Lett.}\ }\textbf {\bibinfo {volume} {61}},\
  \bibinfo {pages} {54} (\bibinfo {year} {1988})}\BibitemShut {NoStop}%
\bibitem [{\citenamefont {Eckert}\ \emph {et~al.}(2002)\citenamefont {Eckert},
  \citenamefont {Schliemann}, \citenamefont {Bruß},\ and\ \citenamefont
  {Lewenstein}}]{ECKERT200288}%
  \BibitemOpen
  \bibfield  {author} {\bibinfo {author} {\bibfnamefont {K.}~\bibnamefont
  {Eckert}}, \bibinfo {author} {\bibfnamefont {J.}~\bibnamefont {Schliemann}},
  \bibinfo {author} {\bibfnamefont {D.}~\bibnamefont {Bruß}},\ and\ \bibinfo
  {author} {\bibfnamefont {M.}~\bibnamefont {Lewenstein}},\ }\href
  {https://doi.org/https://doi.org/10.1006/aphy.2002.6268} {\bibfield
  {journal} {\bibinfo  {journal} {Annals of Physics}\ }\textbf {\bibinfo
  {volume} {299}},\ \bibinfo {pages} {88} (\bibinfo {year} {2002})}\BibitemShut
  {NoStop}%
\bibitem [{\citenamefont {Birrell}\ and\ \citenamefont
  {Davies}(1982)}]{birrell1984quantum}%
  \BibitemOpen
  \bibfield  {author} {\bibinfo {author} {\bibfnamefont {N.}~\bibnamefont
  {Birrell}}\ and\ \bibinfo {author} {\bibfnamefont {P.~C.}\ \bibnamefont
  {Davies}},\ }\href {https://doi.org/10.1017/CBO9780511622632} {\emph
  {\bibinfo {title} {Quantum Fields in Curved Space}}},\ Cambridge Monographs
  on Mathematical Physics\ (\bibinfo  {publisher} {Cambridge University
  Press},\ \bibinfo {year} {1982})\BibitemShut {NoStop}%
\bibitem [{\citenamefont {Exirifard}\ \emph {et~al.}(2021)\citenamefont
  {Exirifard}, \citenamefont {Culf},\ and\ \citenamefont
  {Karimi}}]{Exirifard:Culf:2021}%
  \BibitemOpen
  \bibfield  {author} {\bibinfo {author} {\bibfnamefont {Q.}~\bibnamefont
  {Exirifard}}, \bibinfo {author} {\bibfnamefont {E.}~\bibnamefont {Culf}},\
  and\ \bibinfo {author} {\bibfnamefont {E.}~\bibnamefont {Karimi}},\
  }\bibfield  {journal} {\bibinfo  {journal} {Communications Physics}\ }\textbf
  {\bibinfo {volume} {4}},\ \href {https://doi.org/10.1038/s42005-021-00671-8}
  {10.1038/s42005-021-00671-8} (\bibinfo {year} {2021})\BibitemShut {NoStop}%
\bibitem [{\citenamefont {Exirifard}\ and\ \citenamefont
  {Karimi}(2021)}]{Exirifard:Karimi:2021}%
  \BibitemOpen
  \bibfield  {author} {\bibinfo {author} {\bibfnamefont {Q.}~\bibnamefont
  {Exirifard}}\ and\ \bibinfo {author} {\bibfnamefont {E.}~\bibnamefont
  {Karimi}},\ }\href@noop {} {\bibinfo {title} {Gravitational distortion on
  photon state at the vicinity of the earth}} (\bibinfo {year} {2021}),\
  \Eprint {https://arxiv.org/abs/2110.13990} {arXiv:2110.13990 [gr-qc]}
  \BibitemShut {NoStop}%
\bibitem [{\citenamefont {Philipp}\ \emph
  {et~al.}(2017{\natexlab{a}})\citenamefont {Philipp}, \citenamefont {Perlick},
  \citenamefont {Puetzfeld}, \citenamefont {Hackmann},\ and\ \citenamefont
  {L\"ammerzahl}}]{PhysRevD.95.104037}%
  \BibitemOpen
  \bibfield  {author} {\bibinfo {author} {\bibfnamefont {D.}~\bibnamefont
  {Philipp}}, \bibinfo {author} {\bibfnamefont {V.}~\bibnamefont {Perlick}},
  \bibinfo {author} {\bibfnamefont {D.}~\bibnamefont {Puetzfeld}}, \bibinfo
  {author} {\bibfnamefont {E.}~\bibnamefont {Hackmann}},\ and\ \bibinfo
  {author} {\bibfnamefont {C.}~\bibnamefont {L\"ammerzahl}},\ }\href
  {https://doi.org/10.1103/PhysRevD.95.104037} {\bibfield  {journal} {\bibinfo
  {journal} {Phys. Rev. D}\ }\textbf {\bibinfo {volume} {95}},\ \bibinfo
  {pages} {104037} (\bibinfo {year} {2017}{\natexlab{a}})}\BibitemShut
  {NoStop}%
\bibitem [{\citenamefont {{Carroll}}\ and\ \citenamefont
  {{Traschen}}(2005)}]{carroll2019spacetime}%
  \BibitemOpen
  \bibfield  {author} {\bibinfo {author} {\bibfnamefont {S.~M.}\ \bibnamefont
  {{Carroll}}}\ and\ \bibinfo {author} {\bibfnamefont {J.}~\bibnamefont
  {{Traschen}}},\ }\href {https://doi.org/10.1063/1.1881902} {\bibfield
  {journal} {\bibinfo  {journal} {Phys. To.}\ }\textbf {\bibinfo {volume}
  {58}},\ \bibinfo {pages} {52} (\bibinfo {year} {2005})}\BibitemShut {NoStop}%
\bibitem [{\citenamefont {Wald}(2010)}]{wald2010general}%
  \BibitemOpen
  \bibfield  {author} {\bibinfo {author} {\bibfnamefont {R.}~\bibnamefont
  {Wald}},\ }\href@noop {} {\emph {\bibinfo {title} {General relativity}}}\
  (\bibinfo  {publisher} {University of Chicago press},\ \bibinfo {year}
  {2010})\BibitemShut {NoStop}%
\bibitem [{\citenamefont {Restuccia}\ \emph {et~al.}(2019)\citenamefont
  {Restuccia}, \citenamefont {Toro{\v{s}}}, \citenamefont {Gibson},
  \citenamefont {Ulbricht}, \citenamefont {Faccio},\ and\ \citenamefont
  {Padgett}}]{restuccia2019photon}%
  \BibitemOpen
  \bibfield  {author} {\bibinfo {author} {\bibfnamefont {S.}~\bibnamefont
  {Restuccia}}, \bibinfo {author} {\bibfnamefont {M.}~\bibnamefont
  {Toro{\v{s}}}}, \bibinfo {author} {\bibfnamefont {G.~M.}\ \bibnamefont
  {Gibson}}, \bibinfo {author} {\bibfnamefont {H.}~\bibnamefont {Ulbricht}},
  \bibinfo {author} {\bibfnamefont {D.}~\bibnamefont {Faccio}},\ and\ \bibinfo
  {author} {\bibfnamefont {M.}~\bibnamefont {Padgett}},\ }\href@noop {}
  {\bibfield  {journal} {\bibinfo  {journal} {Phys. Rev. Lett.}\ }\textbf
  {\bibinfo {volume} {123}},\ \bibinfo {pages} {110401} (\bibinfo {year}
  {2019})}\BibitemShut {NoStop}%
\bibitem [{\citenamefont {Brady}\ and\ \citenamefont {Haldar}(2021)}]{Sagnac}%
  \BibitemOpen
  \bibfield  {author} {\bibinfo {author} {\bibfnamefont {A.}~\bibnamefont
  {Brady}}\ and\ \bibinfo {author} {\bibfnamefont {S.}~\bibnamefont {Haldar}},\
  }\href {https://doi.org/10.1103/PhysRevResearch.3.023024} {\bibfield
  {journal} {\bibinfo  {journal} {Phys. Rev. Res.}\ }\textbf {\bibinfo {volume}
  {3}},\ \bibinfo {pages} {023024} (\bibinfo {year} {2021})}\BibitemShut
  {NoStop}%
\bibitem [{\citenamefont {Ramelow}\ \emph {et~al.}(2009)\citenamefont
  {Ramelow}, \citenamefont {Ratschbacher}, \citenamefont {Fedrizzi},
  \citenamefont {Langford},\ and\ \citenamefont {Zeilinger}}]{Zeillinger}%
  \BibitemOpen
  \bibfield  {author} {\bibinfo {author} {\bibfnamefont {S.}~\bibnamefont
  {Ramelow}}, \bibinfo {author} {\bibfnamefont {L.}~\bibnamefont
  {Ratschbacher}}, \bibinfo {author} {\bibfnamefont {A.}~\bibnamefont
  {Fedrizzi}}, \bibinfo {author} {\bibfnamefont {N.}~\bibnamefont {Langford}},\
  and\ \bibinfo {author} {\bibfnamefont {A.}~\bibnamefont {Zeilinger}},\ }\href
  {https://doi.org/10.1103/PhysRevLett.103.253601} {\bibfield  {journal}
  {\bibinfo  {journal} {Phys. Rev. Lett.}\ }\textbf {\bibinfo {volume} {103}},\
  \bibinfo {pages} {253601} (\bibinfo {year} {2009})}\BibitemShut {NoStop}%
\bibitem [{\citenamefont {Fedrizzi}\ \emph {et~al.}(2009)\citenamefont
  {Fedrizzi}, \citenamefont {Herbst}, \citenamefont {Aspelmeyer}, \citenamefont
  {Barbieri}, \citenamefont {Jennewein},\ and\ \citenamefont
  {Zeilinger}}]{Zeillinger2}%
  \BibitemOpen
  \bibfield  {author} {\bibinfo {author} {\bibfnamefont {A.}~\bibnamefont
  {Fedrizzi}}, \bibinfo {author} {\bibfnamefont {T.}~\bibnamefont {Herbst}},
  \bibinfo {author} {\bibfnamefont {M.}~\bibnamefont {Aspelmeyer}}, \bibinfo
  {author} {\bibfnamefont {M.}~\bibnamefont {Barbieri}}, \bibinfo {author}
  {\bibfnamefont {T.}~\bibnamefont {Jennewein}},\ and\ \bibinfo {author}
  {\bibfnamefont {A.}~\bibnamefont {Zeilinger}},\ }\href@noop {} {\bibfield
  {journal} {\bibinfo  {journal} {New J. Phys.}\ }\textbf {\bibinfo {volume}
  {11}},\ \bibinfo {pages} {103052} (\bibinfo {year} {2009})}\BibitemShut
  {NoStop}%
\bibitem [{\citenamefont {Liu}\ \emph {et~al.}(2020)\citenamefont {Liu},
  \citenamefont {Liu}, \citenamefont {Yu},\ and\ \citenamefont
  {Zhang}}]{liu2020sub}%
  \BibitemOpen
  \bibfield  {author} {\bibinfo {author} {\bibfnamefont {J.}~\bibnamefont
  {Liu}}, \bibinfo {author} {\bibfnamefont {J.}~\bibnamefont {Liu}}, \bibinfo
  {author} {\bibfnamefont {P.}~\bibnamefont {Yu}},\ and\ \bibinfo {author}
  {\bibfnamefont {G.}~\bibnamefont {Zhang}},\ }\href@noop {} {\bibfield
  {journal} {\bibinfo  {journal} {APL Photonics}\ }\textbf {\bibinfo {volume}
  {5}},\ \bibinfo {pages} {066105} (\bibinfo {year} {2020})}\BibitemShut
  {NoStop}%
\bibitem [{\citenamefont {Philipp}\ \emph
  {et~al.}(2017{\natexlab{b}})\citenamefont {Philipp}, \citenamefont {Perlick},
  \citenamefont {Puetzfeld}, \citenamefont {Hackmann},\ and\ \citenamefont
  {L\"ammerzahl}}]{philipp2017definition}%
  \BibitemOpen
  \bibfield  {author} {\bibinfo {author} {\bibfnamefont {D.}~\bibnamefont
  {Philipp}}, \bibinfo {author} {\bibfnamefont {V.}~\bibnamefont {Perlick}},
  \bibinfo {author} {\bibfnamefont {D.}~\bibnamefont {Puetzfeld}}, \bibinfo
  {author} {\bibfnamefont {E.}~\bibnamefont {Hackmann}},\ and\ \bibinfo
  {author} {\bibfnamefont {C.}~\bibnamefont {L\"ammerzahl}},\ }\href
  {https://doi.org/10.1103/PhysRevD.95.104037} {\bibfield  {journal} {\bibinfo
  {journal} {Phys. Rev. D}\ }\textbf {\bibinfo {volume} {95}},\ \bibinfo
  {pages} {104037} (\bibinfo {year} {2017}{\natexlab{b}})}\BibitemShut
  {NoStop}%
\bibitem [{\citenamefont {Chen}\ \emph {et~al.}(2019)\citenamefont {Chen},
  \citenamefont {Fink}, \citenamefont {Steinlechner}, \citenamefont {Torres},\
  and\ \citenamefont {Ursin}}]{chen2019hong}%
  \BibitemOpen
  \bibfield  {author} {\bibinfo {author} {\bibfnamefont {Y.}~\bibnamefont
  {Chen}}, \bibinfo {author} {\bibfnamefont {M.}~\bibnamefont {Fink}}, \bibinfo
  {author} {\bibfnamefont {F.}~\bibnamefont {Steinlechner}}, \bibinfo {author}
  {\bibfnamefont {J.~P.}\ \bibnamefont {Torres}},\ and\ \bibinfo {author}
  {\bibfnamefont {R.}~\bibnamefont {Ursin}},\ }\href@noop {} {\bibfield
  {journal} {\bibinfo  {journal} {npj Quantum Information}\ }\textbf {\bibinfo
  {volume} {5}},\ \bibinfo {pages} {1} (\bibinfo {year} {2019})}\BibitemShut
  {NoStop}%
\bibitem [{\citenamefont {Bruschi}\ \emph {et~al.}(2021)\citenamefont
  {Bruschi}, \citenamefont {Chatzinotas}, \citenamefont {Wilhelm},\ and\
  \citenamefont {Schell}}]{Bruschi3}%
  \BibitemOpen
  \bibfield  {author} {\bibinfo {author} {\bibfnamefont {D.}~\bibnamefont
  {Bruschi}}, \bibinfo {author} {\bibfnamefont {S.}~\bibnamefont
  {Chatzinotas}}, \bibinfo {author} {\bibfnamefont {F.}~\bibnamefont
  {Wilhelm}},\ and\ \bibinfo {author} {\bibfnamefont {A.}~\bibnamefont
  {Schell}},\ }\href {https://doi.org/10.1103/PhysRevD.104.085015} {\bibfield
  {journal} {\bibinfo  {journal} {Phys. Rev. D}\ }\textbf {\bibinfo {volume}
  {104}},\ \bibinfo {pages} {085015} (\bibinfo {year} {2021})}\BibitemShut
  {NoStop}%
\bibitem [{\citenamefont {Zych}\ \emph {et~al.}(2011)\citenamefont {Zych},
  \citenamefont {Costa}, \citenamefont {Pikovski},\ and\ \citenamefont
  {Brukner}}]{zych2011quantum}%
  \BibitemOpen
  \bibfield  {author} {\bibinfo {author} {\bibfnamefont {M.}~\bibnamefont
  {Zych}}, \bibinfo {author} {\bibfnamefont {F.}~\bibnamefont {Costa}},
  \bibinfo {author} {\bibfnamefont {I.}~\bibnamefont {Pikovski}},\ and\
  \bibinfo {author} {\bibfnamefont {C.}~\bibnamefont {Brukner}},\ }\href@noop
  {} {\bibfield  {journal} {\bibinfo  {journal} {Nature communications}\
  }\textbf {\bibinfo {volume} {2}},\ \bibinfo {pages} {1} (\bibinfo {year}
  {2011})}\BibitemShut {NoStop}%
\end{thebibliography}%

\appendix

\end{document}